\documentclass[twocolumn,aps,pra,floatfix,superscriptaddress,notitlepage]{revtex4-1}
\usepackage[utf8]{inputenc}
\usepackage[OT1]{fontenc}
\usepackage{amsmath}
\usepackage{amsfonts}
\usepackage{amssymb}
\usepackage{bm}
\usepackage{graphicx}
\usepackage[left=2cm,right=2cm,top=2cm,bottom=2cm]{geometry}
\usepackage{booktabs}
\allowdisplaybreaks

\usepackage{dcolumn}
\usepackage{siunitx}
\usepackage{multirow}
\usepackage[caption=false]{subfig}

\usepackage{color}
\definecolor{BLUE}{rgb}{0.0,0.0,1.0}

\usepackage{soul}
\soulregister\cite7
\soulregister\ref7

\newcommand{\veps}{\varepsilon}
\newcommand{\balpha}{\bm{\alpha}}
\newcommand{\bnabla}{\bm{\nabla}}

\newcommand{\bD}{\bm{D}}
\newcommand{\br}{\bm{r}}

\newcommand{\bp}{\bm{p}}

\newcommand{\be}{\begin{eqnarray}}
\newcommand{\ee}{\end{eqnarray}}

\newcommand{\matr}[3]{\langle #1 | #2 | #3 \rangle}

\begin{document}

\title{QED theory of the normal mass shift in few-electron atoms}

%

\author{A.~V.~Malyshev}
\affiliation{Department of Physics, St.~Petersburg State University, Universitetskaya 7/9, 199034 St.~Petersburg, Russia  
\looseness=-1}

\author{I.~S.~Anisimova}
\affiliation{Department of Physics, St.~Petersburg State University, Universitetskaya 7/9, 199034 St.~Petersburg, Russia  
\looseness=-1}

\author{D.~A.~Glazov}
\affiliation{Department of Physics, St.~Petersburg State University, Universitetskaya 7/9, 199034 St.~Petersburg, Russia  
\looseness=-1}

\author{M.~Y.~Kaygorodov}
\affiliation{Department of Physics, St.~Petersburg State University, Universitetskaya 7/9, 199034 St.~Petersburg, Russia  
\looseness=-1}

\author{D.~V.~Mironova}
\affiliation{St.~Petersburg Electrotechnical University, Prof. Popov 5, 197376 St.~Petersburg, Russia
\looseness=-1}

\author{G.~Plunien}
\affiliation{Institut f\"ur Theoretische Physik, Technische Universit\"at Dresden, Mommsenstra{\ss}e 13, D-01062 Dresden, Germany
\looseness=-1}

\author{V.~M.~Shabaev}
\affiliation{Department of Physics, St.~Petersburg State University, Universitetskaya 7/9, 199034 St.~Petersburg, Russia  
\looseness=-1}


\begin{abstract}

The electron-electron interaction correction of first order in $1/Z$ to the one-electron part of the nuclear recoil effect on binding energies in atoms and ions is considered within the framework of the rigorous QED approach. The calculations to all orders in $\alpha Z$ are performed for the $1s^2$ state in heliumlike ions and the $1s^2 2s$ and $1s^2 2p_{1/2}$ states in lithiumlike ions in the range $Z=5$--$100$. The results obtained are compared with the Breit-approximation values. The performed calculations complete a systematic treatment of the QED nuclear recoil effect up to the first order in $1/Z$. The correction obtained is combined with the previously studied two-electron part as well as the higher-order electron-correlation corrections evaluated within the Breit approximation to get the total theoretical predictions for the mass shifts.  

\end{abstract}


\maketitle


\section{Introduction \label{sec:0}}

It is well known that within the nonrelativistic approximation the effect of the nuclear motion on spectra of hydrogenlike ions is accounted for exactly by replacing the electron mass $m$ with the reduced mass $m_r=mM/(m+M)$ with $M$ being the mass of the nucleus. The lowest-order relativistic correction of first order in $m/M$ can be derived from the Breit equation for electron and nucleus~\cite{Bethe:1957:book}. For $N$-electron system, the corresponding one-electron contribution to the nuclear recoil effect can be described by the operator 
\begin{align}
\label{eq:NMS}
H_{\rm NMS} = \frac{1}{2M} \sum_{i}^{N} \left\{ 
        \bp_i^2
      - \frac{\alpha Z}{r_i} \left[ \balpha_i + \frac{(\balpha_i \cdot \br_i) \br_i }{r_i^2} \right] \cdot \bp_i 
                             \right\} \, ,
\end{align}  
where $\bp=-i\bnabla$ is the momentum operator, $\br$ is the position vector, $r=|\br|$, $\balpha$ are the Dirac matrices, $\alpha$ is the fine-structure constant, and $Z$ is the nuclear charge number [the relativistic units ($\hbar=1$, $c=1$) are used throughout the paper]. The one-electron operator in Eq.~(\ref{eq:NMS}) gives rise to the so-called normal mass shift (NMS). In the case of more than one electron, the NMS operator~(\ref{eq:NMS}) does not provide the exhaustive description of the effect of the nuclear motion since there is also the two-electron contribution given by the specific mass shift (SMS) operator
\begin{align}
\label{eq:SMS}
H_{\rm SMS} = \frac{1}{2M} \sum_{i\neq j}^{N} \left\{ 
        \bp_i \cdot \bp_j  
      - \frac{\alpha Z}{r_i} \left[ \balpha_i + \frac{(\balpha_i \cdot \br_i) \br_i }{r_i^2} \right] \cdot \bp_j 
                             \right\} \, .
\end{align} 
The NMS and SMS operators add to the mass shift (MS) operator~\cite{Shabaev:1985:588, Shabaev:1988:69, Palmer:1987:5987},
\begin{align}
\label{eq:NMS+SMS}
H_{M}= H_{\rm NMS} + H_{\rm SMS} \, , 
\end{align} 
which allows one to treat the nuclear recoil contribution within the $(m/M)(\alpha Z)^4mc^2$ approximation. To date, the MS operator~(\ref{eq:NMS+SMS}) is used extensively in relativistic calculations of the atomic spectra and isotope shifts (see, e.g., Refs.~\cite{Tupitsyn:2003:022511, SoriaOrts:2006:103002, Korol:2007:022103, Kozhedub:2010:042513, Gaidamauskas:2011:175003, Naze:2014:1197, Zubova:2014:062512, Naze:2015:032511, Zubova:2016:052502, Filippin:2017:042502, Tupitsyn:2018:022517, Gamrath:2018:38, Zubova:2019:185001, Ekman:2019:433, Zaytsev:2019:052504, Yerokhin:2020:012502} and references therein).

The fully relativistic description of the nuclear recoil effect on binding energies requires application of the bound-state quantum electrodynamics (QED) beyond the Breit approximation. The corresponding theory to first order in $m/M$ and to all orders in $\alpha Z$ was developed in Refs.~\cite{Shabaev:1985:588, Shabaev:1988:69, Shabaev:1998:59}; see also Refs.~\cite{Pachucki:1995:1854, Yelkhovsky:Budker, Adkins:2007:042508}. Numerous QED evaluations of the nuclear recoil contribution to binding energies were performed over the past three decades~\cite{Artemyev:1995:1884, Artemyev:1995:5201, Shabaev:1998:4235, Shabaev:1999:493, SoriaOrts:2006:103002, Adkins:2007:042508, Zubova:2016:052502, Malyshev:2018:085001}. However, all the previous nonperturbative (in $\alpha Z$) calculations were limited by the independent-electron approximation, i.e., the electron-electron interaction corrections to the nuclear recoil effect were neglected. It should be noted that the interelectronic-interaction effects were treated approximately in some cases by modifying the zeroth-order approximation and including into it a local screening potential (see, e.g., Ref.~\cite{SoriaOrts:2006:103002}). In our recent work~\cite{Malyshev:2019:012510}, we addressed the issue of the QED evaluation of the interelectronic-interaction correction of first order in $1/Z$ to the two-electron part of the nuclear recoil effect on binding energies. The present paper focuses on deriving the rigorous QED formalism for calculations of the corresponding correction to the dominant one-electron part. The results obtained represent the nontrivial QED contribution to the NMS and complete the rigorous consideration of the first-order (in $1/Z$) nuclear recoil effect to all orders in $\alpha Z$. 

The QED formalism worked out in the present work is illustrated by calculating the one-electron part of the nuclear recoil effect on binding energies of the $1s^2$ state in heliumlike ions and the $1s^2 2s$ and $1s^2 2p_{1/2}$ states in lithiumlike ions for the wide range of the nuclear charge number $Z=5$--$100$. The behavior of the nontrivial QED contribution to the NMS as a function of $Z$ is studied. These calculations together with those performed in Ref.~\cite{Malyshev:2019:012510} provide a better understanding of the applicability limits for the MS operator~(\ref{eq:NMS+SMS}). In particular, one can assume that the application of a rigorous QED approach will resolve some discrepancies which take place nowadays between the preliminary calculations and the high-precision measurements of the isotope shifts of the fine-structure splittings in singly ionized argon (${\rm Ar}^+$)~\cite{Botsi:ICPEAC, Botsi:MS_thesis} and calcium (${\rm Ca}^+$)~\cite{Shi:2016:2}. We also stress that the effect under consideration may contribute significantly when specific differences of the energies or isotope shifts are studied (see, e.g., Ref.~\cite{Malyshev:2017:765} for the related discussion in the case of the bound-electron $g$ factor). 

The paper is organized as follows. The main aspects of the QED theory of the nuclear recoil effect on binding energies within the independent-electron approximation are outlined in Sec.~\ref{sec:1}. The formulas for the first-order (in $1/Z$) correction to the one-electron part of the nuclear recoil effect valid to all orders in $\alpha Z$ are discussed in Sec.~\ref{sec:2}. The numerical results and the comparison with the values obtained employing the MS Hamiltonian~(\ref{eq:NMS+SMS}) are given in Sec.~\ref{sec:3}.


\section{QED theory of the nuclear recoil effect within the independent-electron approximation \label{sec:1}}

The QED theory of the nuclear recoil effect on atomic binding energies was worked out in Refs.~\cite{Shabaev:1985:588, Shabaev:1988:69, Shabaev:1998:59}. The formulation of the theory presented in Ref.~\cite{Shabaev:1998:59} is the most convenient for the needs of the present study. It reduces the problem of accounting for the nuclear recoil effect to a modification of the standard QED Hamiltonian of the electron-positron field interacting with the quantized electromagnetic field and the classical Coulomb potential of the nucleus $V_{\rm n}$. The modification consists in an extra term to the interaction part of the QED Hamiltonian; see Ref.~\cite{Shabaev:1998:59} for the details. As a result, the nuclear recoil effect to first order in $m/M$ and to all orders in $\alpha Z$ can be taken into account by perturbation theory in the interaction representation of the Furry picture~\cite{Furry:1951:115}. For the construction of the perturbation series, we employ the two-time Green's function (TTGF) method~\cite{TTGF}. All the necessary Feynman rules can be found, e.g., in Ref.~\cite{TTGF}. In order to describe the new elements of the diagram technique as compared to the standard bound-state QED, we discuss briefly the derivation of the formulas for the one-electron part of the nuclear recoil effect to zeroth order in $1/Z$ for the electron in the state $|a\rangle$. The total one-electron contribution for a given many-electron state is obtained by adding the corresponding terms from all one-electron orbitals.

\begin{figure}
\begin{center}
\includegraphics[width=0.9\columnwidth]{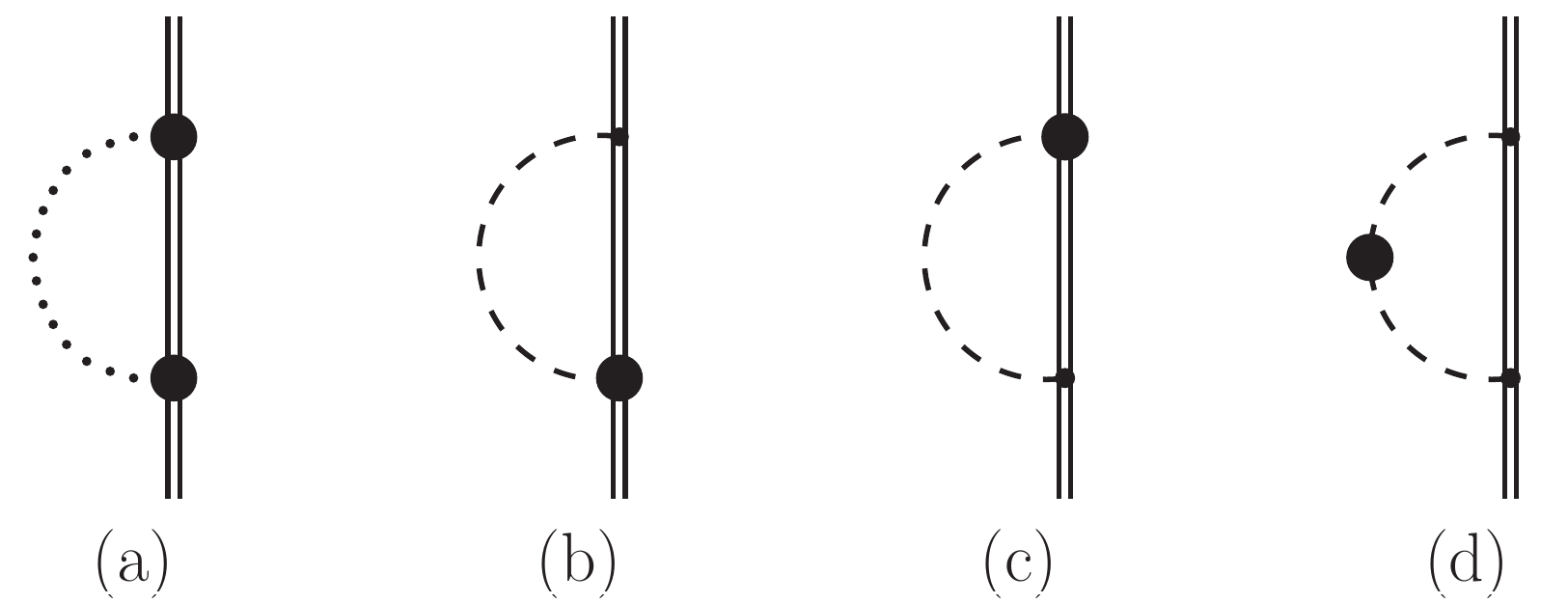}
\caption{\label{fig:recoil_1el}
One-electron nuclear recoil diagrams to zeroth order in $1/Z$: the Coulomb~(a), one-transverse~(b) and (c), and two-transverse~(d) contributions. See the text and Ref.~\cite{Shabaev:1998:59} for the description of the Feynman rules.}
\end{center}
\end{figure}

The one-electron part of the nuclear recoil effect is given by the diagrams depicted in Fig.~\ref{fig:recoil_1el}. The double line denotes the propagator for an electron in the classical field of the nucleus. The vertex with a small black dot corresponds to the conventional QED vertex. The new vertex with a bold dot arises from the extra term to the QED Hamiltonian derived in Ref.~\cite{Shabaev:1998:59}. It contains the momentum operator $\bp$. Following the notations employed in Ref.~\cite{Shabaev:1998:59}, we refer to the dotted line joining two bold dots in Fig.~\ref{fig:recoil_1el}(a) as to the ``Coulomb recoil'' interaction. The dashed line ended by a bold dot on one side in Figs.~\ref{fig:recoil_1el}(b) and \ref{fig:recoil_1el}(c) designates the ``one-transverse-photon recoil'' interaction, since it includes the transverse part of the photon propagator in the Coulomb gauge
\begin{align}
\label{eq:D_lk}
D_{lk}(\omega,\br) &= 
-\frac{1}{4\pi} \Bigg[ 
\frac{\exp\left( i \sqrt{\omega^2 + i0} \, r \right)}{ r } \delta_{lk}    \nonumber \\
&\qquad\quad
+
\nabla_l \nabla_k
\frac{\exp\left( i \sqrt{\omega^2 + i0} \, r \right) - 1}{\omega^2 r}
\Bigg] \, ,
\end{align}
where we fix the branch of the square root by the condition ${\rm Im}\left(\sqrt{\omega^2 + i0}\right)>0$. The dashed line with a bold dot on it in Fig.~\ref{fig:recoil_1el}(d) corresponds to the ``two-transverse-photon recoil'' interaction, since it involves the product of two photon propagators (\ref{eq:D_lk}). The terminology used comes from operating in the Coulomb gauge which appears to be the most appropriate and convenient gauge for studying the nuclear recoil effect; see, e.g., Refs.~\cite{Shabaev:1985:588, Shabaev:1988:69, Yelkhovsky:Budker}.

Within the Furry picture, the zeroth-order approximation for one-electron energies and wave functions is determined by the Dirac equation with the binding potential of the nucleus $V_{\rm n}$,
\begin{equation}
\label{eq:DirEq}
\left[ -i\balpha \cdot \bnabla + \beta m +V_{\rm n}(r) \right] \psi_n(\br) = \veps_n \psi_n(\br) \, .
\end{equation}
The TTGF method prescribes that the first-order correction to the energy of an arbitrary single level $|u\rangle$ can be obtained according to the following formula:
\begin{align}
\label{eq:dE1}
\Delta E^{(1)}  
=
\frac{1}{2\pi i} \oint_\Gamma \! dE \, \Delta E \Delta g^{(1)}_{uu}(E) \, .
\end{align}
Here $\Delta g^{(1)}_{uu}$ represents the Fourier transform of the contribution to the two-time Green's function projected on the unperturbed state~$|u^{(0)}\rangle$, $\Delta E = E-E_{u}^{(0)}$, $E_{u}^{(0)}$ is the unperturbed energy, and the oriented counterclockwise contour $\Gamma$ surrounds $E_{u}^{(0)}$ in the complex $E$ plane; see Ref.~\cite{TTGF} for the details. For the one-electron nuclear recoil contribution under consideration, we assume that the unperturbed wave function $|u_{\rm 1el}^{(0)}\rangle$ is given by the solution~$\psi_a$ of the Dirac equation~(\ref{eq:DirEq}), and the unperturbed energy coincides with the corresponding energy~$\veps_a$. The diagrams shown in Fig.~\ref{fig:recoil_1el} are similar to the first-order self-energy diagram; see, e.g., Refs.~\cite{Mohr:1974:26, Mohr:1974:52, Yerokhin:1999:800}. For this reason, the derivation of the formulas does not cause any problems, and by applying the TTGF method one readily obtains
\begin{align}
\label{eq:dE_c_1el}
\Delta E^{(1)}_{\rm c} = 
\frac{1}{M} \frac{i}{2\pi} 
\int\limits_{-\infty}^{\infty} \! d\omega \, \sum_n
\frac{ \matr{a}{p_k}{n} \matr{n}{p_k}{a} }
     { \omega + \veps_a - u\veps_n } 
\end{align}
for the Coulomb contribution in Fig.~\ref{fig:recoil_1el}(a),
\begin{align}
\label{eq:dE_tr1_1el}
\Delta E^{(1)}_{\rm tr1} = 
-\frac{1}{M} \frac{i}{2\pi} 
\int\limits_{-\infty}^{\infty} \! d\omega \, \sum_n
\Bigg[
\frac{ \matr{a}{p_k}{n} \matr{n}{D_k(\omega)}{a} }
     { \omega + \veps_a - u\veps_n }       \nonumber \\ 
 + 
\frac{ \matr{a}{D_k(\omega)}{n} \matr{n}{p_k}{a} }
     { \omega + \veps_a - u\veps_n }   
\Bigg]  
\end{align}
for the one-transverse-photon contribution in Figs.~\ref{fig:recoil_1el}(b) and \ref{fig:recoil_1el}(c), and
\begin{align}
\label{eq:dE_tr2_1el}
\Delta E^{(1)}_{\rm tr2} = 
\frac{1}{M} \frac{i}{2\pi} 
\int\limits_{-\infty}^{\infty} \! d\omega \, \sum_n
\frac{ \matr{a}{D_k(\omega)}{n} \matr{n}{D_k(\omega)}{a} }
     { \omega + \veps_a - u\veps_n } 
\end{align}
for the two-transverse-photon contribution in Fig.~\ref{fig:recoil_1el}(d). In Eqs.~(\ref{eq:dE_c_1el})-(\ref{eq:dE_tr2_1el}) and below,
the summation over the repeated indices is implied, $u=1-i0$ provides the proper treatment of the poles in the electron propagator, and
\begin{align}
\label{eq:D}
D_k(\omega) = -4\pi \alpha Z \alpha_l D_{lk} (\omega) \, ,
\end{align}
where $\alpha_l$ $(l=1,2,3)$ are the Dirac matrices. For the following, it is convenient to introduce the notations
\begin{align}
\label{eq:R_c}
R_{\rm c} &= \frac{1}{M} \, \bp_1 \cdot \bp_2 \, ,     \\
\label{eq:R_tr1}
R_{\rm tr1}(\omega) &= -\frac{1}{M} \, \big[ \bp_1 \cdot \bD_2(\omega) + \bD_1(\omega) \cdot \bp_2 \big] \, ,  \\
\label{eq:R_tr2}
R_{\rm tr2}(\omega) &= \frac{1}{M} \, \bD_1(\omega) \cdot \bD_2(\omega) \, ,   
\end{align}
for the Coulomb, one-transverse-photon, and two-transverse-photon interactions, respectively. By analogy with the self-energy operator $\Sigma(E)$, we also introduce the operator ${\rm P}(E)$ for the nuclear recoil effect,
\begin{align}
\label{eq:P_operator}
\langle a | {\rm P}(E) | b \rangle
=
\frac{i}{2\pi} \int\limits_{-\infty}^{\infty} \! d\omega \, \sum_n 
\frac{\langle a n | R(\omega) | n b \rangle}{\omega + E - u\veps_n} \, ,
\end{align}
where $R$ means any of the operators~(\ref{eq:R_c})--(\ref{eq:R_tr2}). Then, Eqs.~(\ref{eq:dE_c_1el})--(\ref{eq:dE_tr2_1el}) can be rewritten as $\langle a | {\rm P}_{\rm c}(\veps_a) | a \rangle$, $\langle a | {\rm P}_{\rm tr1}(\veps_a) | a \rangle$, and $\langle a | {\rm P}_{\rm tr2}(\veps_a) | a \rangle$, respectively. Finally, the total one-electron contribution to the nuclear recoil effect to zeroth order in $1/Z$ is given by the sum of Eqs.~(\ref{eq:dE_c_1el})--(\ref{eq:dE_tr2_1el}),
\begin{align}
\label{eq:dE_rec_1el}
\Delta E^{(1)}_{\rm 1el} = \Delta E^{(1)}_{\rm c} + \Delta E^{(1)}_{\rm tr1} + \Delta E^{(1)}_{\rm tr2} \, .
\end{align}

The integration over $\omega$ in the Coulomb contribution~(\ref{eq:dE_c_1el}) can be evaluated analytically using the standard identity ($\omega_1<0<\omega_2$):
\begin{align}
\label{eq:sokhot}
\int\limits_{\omega_1}^{\omega_2} \! d\omega \, \frac{f(\omega)}{\omega \pm i0} =
\mp i\pi f(0) + {\mathcal{P}} \int\limits_{\omega_1}^{\omega_2} \! d\omega \, \frac{f(\omega)}{\omega} \, , 
\end{align}
where ${\mathcal{P}}$ means the principal value integral. Indeed, applying the formula~(\ref{eq:sokhot}) to Eq.~(\ref{eq:dE_c_1el}) and taking into account that all the principal value integrals vanish, one obtains
\begin{align}
\label{eq:dE_c_1el_calc}
\Delta E^{(1)}_{\rm c} &= 
\frac{1}{2M}
\sum_n^{\veps_n>0} \matr{a}{p_k}{n} \matr{n}{p_k}{a}   \nonumber \\
&\, -
\frac{1}{2M}
\sum_n^{\veps_n<0} \matr{a}{p_k}{n} \matr{n}{p_k}{a} \, ,
\end{align}
where the first and second summations run over the positive- and negative-energy parts of the spectrum, respectively. It is useful to compare this expression with the formula which can be obtained by employing the nonrelativistic part of the NMS operator (\ref{eq:NMS}):
\begin{align}
\label{eq:dE_c_Breit}
\Delta E^{(1)}_{\rm c,Breit} =
\left\langle a \left| \frac{\bp^2}{2M} \right| a \right\rangle 
=
\frac{1}{2M}
\sum_n \matr{a}{p_k}{n} \matr{n}{p_k}{a} \, ,
\end{align}
where the summation runs over all the states. One can see that introducing the projectors on the positive-energy part of the spectrum in Eqs.~(\ref{eq:dE_c_1el}) or (\ref{eq:dE_c_1el_calc}) leads to the result which differs from the value~(\ref{eq:dE_c_Breit}) by the contribution of the negative-energy continuum, being of order $(m/M)(\alpha Z)^5mc^2$, i.e., beyond the Breit approximation. The expression~(\ref{eq:dE_c_Breit}) is implied to be the lowest-order approximation of the Coulomb contribution~(\ref{eq:dE_c_1el}). One should note that Eq.~(\ref{eq:dE_c_Breit}) contains actually some terms of the higher orders in $\alpha Z$ as well, since it is evaluated with the Dirac wave functions. The nontrivial QED Coulomb contribution, which can not be obtained from the Breit equation, reads~\cite{Artemyev:1995:1884}
\begin{align}
\label{eq:dE_c_QED}
\Delta E^{(1)}_{\rm c,QED} &\equiv \Delta E^{(1)}_{\rm c} - \Delta E^{(1)}_{\rm c,Breit}  \nonumber \\
&=
-\frac{1}{M} \sum_n^{\veps_n<0} \matr{a}{p_k}{n} \matr{n}{p_k}{a} \, .
\end{align}

In order to obtain the lowest-order relativistic approximation to the one-transverse-photon contribution~(\ref{eq:dE_tr1_1el}), one has to consider the zero-energy-transfer limit $\omega\to0$ of Eq.~(\ref{eq:D}) given by
\begin{align}
\label{eq:D0}
D_k(0) = \frac{\alpha Z}{2r} 
\left[
\alpha_k + \frac{(\alpha_i r_i) r_k}{r^2}
\right] \, .
\end{align}
By neglecting the energy dependence of the vector $\bD(\omega)$ in Eq.~(\ref{eq:dE_tr1_1el}), we come to the integral which is similar to the Coulomb case~(\ref{eq:dE_c_1el}). One should take care defining its Breit approximation, since discarding the negative-energy part of the spectrum in Eq.~(\ref{eq:dE_tr1_1el}) leads once again to a slightly different result. As in the Coulomb case, we consider the expression arising from the NMS operator,  
\begin{align}
\Delta E^{(1)}_{\rm tr1,Breit} = -\frac{1}{2M} \matr{a}{(\bp\cdot\bD(0)+\bD(0)\cdot\bp)}{a} \, ,
\end{align}
as the lowest-order relativistic approximation to Eq.~(\ref{eq:dE_tr1_1el}), and the nontrivial QED part of the one-transverse-photon contribution is
\begin{align}
\label{eq:dE_tr1_QED}
\Delta E^{(1)}_{\rm tr1,QED} &\equiv \Delta E^{(1)}_{\rm tr1} - \Delta E^{(1)}_{\rm tr1,Breit} \, .
\end{align}
We note, finally, that the two-transverse-photon contribution~$\Delta E^{(1)}_{\rm tr2}$ is completely beyond the Breit approximation. Thus, we relegate it to the nontrivial QED part. 

The total one-electron nuclear recoil contribution (\ref{eq:dE_rec_1el}) can be conveniently represented as a sum of the Breit-approximation term and the nontrivial QED term,
\begin{align}
\label{eq:dE_rec_L+H}
\Delta E^{(1)}_{\rm 1el} &= \Delta E^{(1)}_{\rm 1el,Breit} + \Delta E^{(1)}_{\rm 1el,QED} \, , \\
\label{eq:dE_rec_L}
\Delta E^{(1)}_{\rm 1el,Breit} &= \Delta E^{(1)}_{\rm c,Breit}+\Delta E^{(1)}_{\rm tr1,Breit} \nonumber \\
& \!\!\!\!\!\!\!\!\!\!\!\!\!\!\!
=\frac{1}{2M} \matr{a}{[\bp^2-(\bp\cdot\bD(0)+\bD(0)\cdot\bp)]}{a} \, \\
\label{eq:dE_rec_H}
\Delta E^{(1)}_{\rm 1el,QED} &=
\Delta E^{(1)}_{\rm c,QED}+\Delta E^{(1)}_{\rm tr1,QED}+\Delta E^{(1)}_{\rm tr2} \nonumber \\
& \!\!\!\!\!\!\!\!\!\!\!\!\!\!\!
=\frac{1}{M} \frac{i}{2\pi} 
\int\limits_{-\infty}^{\infty} \! d\omega \, 
\langle a | 
\left(
D_k(\omega) - \frac{[p_k,V_{\rm n}]}{\omega+i0}
\right)  \nonumber \\
&\times
G(\omega+\veps_a)
\left(
D_k(\omega) + \frac{[p_k,V_{\rm n}]}{\omega+i0}
\right)
| a \rangle \, ,
\end{align}
where $G(\omega)=\sum_n|n\rangle \langle n|[\omega-u\veps_n]^{-1}$ is the Dirac-Coulomb Green's function and $[A,B]=AB-BA$. The formalism for treating the nuclear recoil effect to all orders in $\alpha Z$ was initially derived in Ref.~\cite{Shabaev:1985:588} in the form given by the Eqs.~(\ref{eq:dE_rec_L+H})--(\ref{eq:dE_rec_H}). 


\section{Electron-electron interaction correction to the one-electron part of the nuclear recoil effect \label{sec:2}}

\begin{figure}
\begin{center}
\includegraphics[width=\columnwidth]{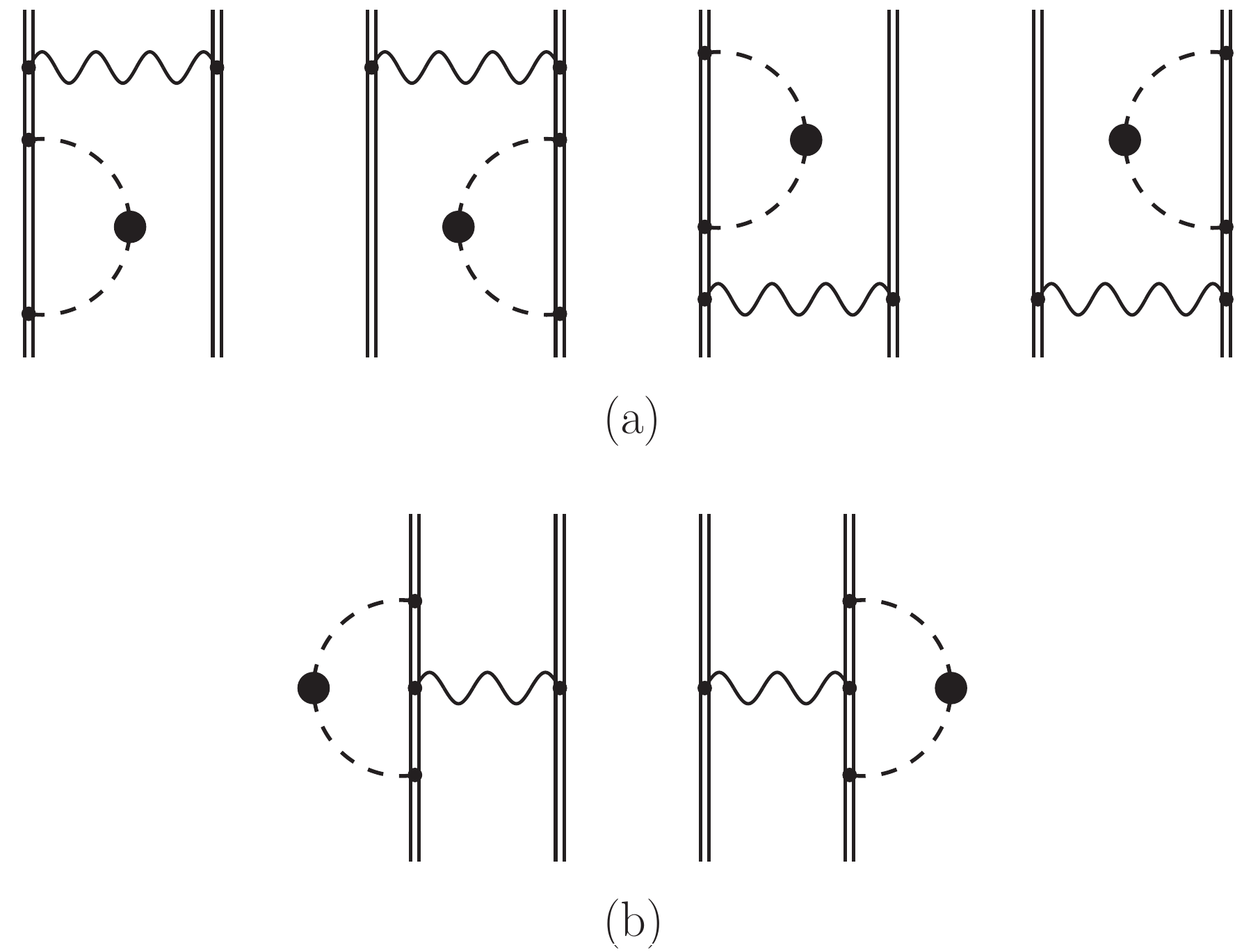}
\caption{\label{fig:recoil_1el_IntEl}
The second-order diagrams describing the electron-electron interaction correction to the one-electron two-transverse-photon contribution to the nuclear recoil effect. The analogous diagrams with the Coulomb and one-transverse photon recoil interactions have to be taken into account as well. See the text and Ref.~\cite{Shabaev:1998:59} for the description of the diagram technique. 
}
\end{center}
\end{figure}

One set of Feynman diagrams contributing to the first order (in $1/Z$) electron-electron interaction correction to the one-electron part of the nuclear recoil effect is shown in Fig.~\ref{fig:recoil_1el_IntEl}. The wavy line corresponds to the photon propagator, while all the other notations are the same as in Fig.~\ref{fig:recoil_1el}. The two-transverse-photon contribution presented in Fig.~\ref{fig:recoil_1el_IntEl} has to be complemented by the corresponding Coulomb and one-transverse-photon contributions. Therefore, the total number of the second-order diagrams is four times higher.

The second-order correction to energy of a single level~$|u\rangle$ is given by \cite{TTGF}
\begin{align}
\label{eq:dE2}
\Delta E^{(2)}  
&= 
\frac{1}{2\pi i} \oint_\Gamma \! dE \, \Delta E \Delta g^{(2)}_{uu}(E) \, \nonumber \\
& \!\!\!\!\!\!\!
-      
\left[
\frac{1}{2\pi i} \oint_\Gamma \! dE \, \Delta E \Delta g^{(1)}_{uu}(E) 
\right]
\left[
\frac{1}{2\pi i} \oint_\Gamma \! dE \, \Delta g^{(1)}_{uu}(E) 
\right] \, ,
\end{align}
where the contour $\Gamma$ surrounds the unperturbed energy~$E^{(0)}_u$ and keeps outside all the other singularities of the Green's function. In this paper we are interested in the two-electron corrections presented in Fig.~\ref{fig:recoil_1el_IntEl}. An arbitrary many-electron problem can be easily decomposed into the set of two-electron problems. For this reason, it is sufficient to assume the unperturbed wave function $|u^{(0)}\rangle$ in Eq.~(\ref{eq:dE2}) to be represented by the one-determinant two-electron wave function,
\begin{align}
\label{eq:u_2el} 
|u_{\rm 2el}^{(0)}\rangle=\frac{1}{\sqrt{2}} \sum_P (-1)^P 
\psi_{Pa} (\br_1) \psi_{Pb} (\br_2)  \, ,
\end{align}
where $\psi_a$ and $\psi_b$ are the solutions of the Dirac equation (\ref{eq:DirEq}), $P$ is the permutation operator, and $(-1)^P$ is the sign of the permutation. The unperturbed energy is given by the sum of the one-electron Dirac energies: $E^{(0)}_u = \veps_a + \veps_b$. The generalization to the case of a many-determinant wave function is straightforward and can be done in the final expressions. 

\begin{figure}
\begin{center}
\includegraphics[width=0.23\columnwidth]{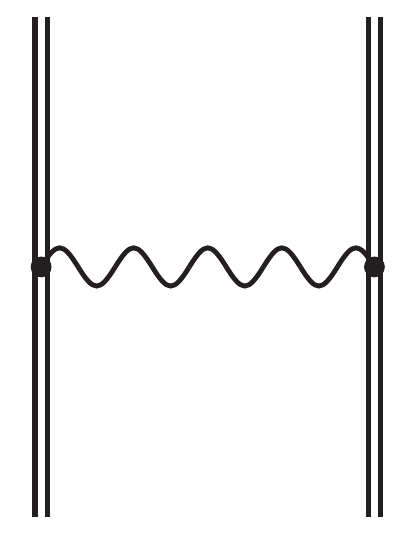}
\caption{\label{fig:1ph}
The one-photon exchange diagram which contributes to the second ``disconnected'' term in Eq.~(\ref{eq:dE2}) along with the first-order diagrams in Fig.~\ref{fig:recoil_1el}. 
}
\end{center}
\end{figure}

The second term in Eq.~(\ref{eq:dE2}), given by the product of the first-order contributions to the Green's function, is usually referred to as the ``disconnected'' one. The relevant diagrams are shown in Figs.~\ref{fig:recoil_1el} and \ref{fig:1ph}. The disconnected term is to be considered together with the related contribution in the first term in Eq.~(\ref{eq:dE2}). As a rule, it is fully canceled analytically by identifying the corresponding expressions. In the following, we will not mention the disconnected term any longer, but its contribution is always taken into account.

Prior to deriving the formulas for the interelectronic-interaction correction to the one-electron part of the nuclear recoil effect, we introduce the operator
\begin{align}
\label{eq:I}
I(\omega) &= e^2 \alpha_1^\mu \alpha_2^\nu D_{\mu\nu}(\omega) \, ,   
\end{align}
where $\alpha^\mu = (1,\balpha)$ and $D_{\mu\nu}$ is the photon propagator. For the Coulomb gauge, in which we operate, Eq.~(\ref{eq:I}) has the form
\begin{align}
\label{eq:I_coul}
I(\omega) &= \alpha \, \Bigg[ \,
\frac{1}{r_{12}} - (\balpha_1\cdot\balpha_2)\frac{\exp\left( i \sqrt{\omega^2 + i0} \, r_{12} \right)}{r_{12}}   
\nonumber \\
&+ 
(\balpha_1\cdot\bnabla_1)
(\balpha_2\cdot\bnabla_2) \,
\frac{\exp\left( i \sqrt{\omega^2 + i0} \, r_{12} \right)-1}{\omega^2 r_{12}} 
\, \Bigg] \, .
\end{align}
The zero-energy-transfer limit $\omega \to 0$ of Eq.~(\ref{eq:I_coul}) reads
\begin{align}
\label{eq:I0}
I(0) &= \alpha \, \Bigg[ \,
\frac{1}{r_{12}} - \frac{(\balpha_1\cdot\balpha_2)}{r_{12}}   
- 
\frac{(\balpha_1\cdot\bnabla_1)(\balpha_2\cdot\bnabla_2)\, r_{12} }{2} 
\, \Bigg] \, .
\end{align}
The operator (\ref{eq:I0}) can be used to evaluate the interelectronic-interaction correction to the MS operator (\ref{eq:NMS+SMS}) within the Breit approximation.

As noted in Sec.~\ref{sec:1}, the diagrams for the one-electron part of the nuclear recoil effect in Fig.~\ref{fig:recoil_1el} are similar to the diagram for the first-order self energy. In turn, the diagrams for the electron-electron interaction correction in Fig.~\ref{fig:recoil_1el_IntEl} are similar to the two-electron self-energy diagrams, which were discussed in details in, e.g., Refs.~\cite{Yerokhin:1999:3522, LeBigot:2001:040501}. For this reason, we present here only the final formulas and omit all the intermediate manipulations. We divide the total interelectronic-interaction correction to the one-electron part of the nuclear recoil effect into three parts. The contribution of the diagrams in Fig.~\ref{fig:recoil_1el_IntEl}(b) is referred to as the ``vertex'' term ($\Delta E_{\rm vert}^{(2)}$). The contribution of the diagrams shown in Fig.~\ref{fig:recoil_1el_IntEl}(a) is naturally divided into the ``irreducible'' ($\Delta E_{\rm irr}^{(2)}$) and ``reducible'' ($\Delta E_{\rm red}^{(2)}$) parts. The reducible part is defined as the contribution in which the energy of the intermediate two-electron state coincides with the energy of the initial state~$E^{(0)}_u$. The irreducible part is the remainder. 

The irreducible contribution reads
\begin{align}
\label{eq:irr_1el}
\Delta E_{\rm irr}^{(2)} =
2 \Big[
\langle a | {\rm P}(\veps_a) | \xi_a \rangle
+
\langle b | {\rm P}(\veps_b) | \xi_b \rangle
\Big] \, ,
\end{align}
where the operator ${\rm P}$ was defined by Eq.~(\ref{eq:P_operator}), 
\begin{align}
\label{eq:xi_a}
| \xi_a \rangle &= 
\sum_{\veps_n\neq\veps_a} \frac{|n\rangle}{\veps_a-\veps_n} \sum_P (-1)^P 
\langle n b | I(\Delta) | Pa Pb \rangle \, ,   \\
\label{eq:xi_b}
| \xi_b \rangle &= 
\sum_{\veps_n\neq\veps_b} \frac{|n\rangle}{\veps_b-\veps_n} \sum_P (-1)^P 
\langle a n | I(\Delta) | Pa Pb \rangle \, , 
\end{align}
and $\Delta = \veps_{Pa} - \veps_a$. The reducible contribution has the form
\begin{align}
\label{eq:red_1el}
\Delta E_{\rm red}^{(2)} &=
\matr{ba}{I'(\Delta)}{ab} 
\Big[
\langle a | {\rm P}(\veps_a) | a \rangle
-
\langle b | {\rm P}(\veps_b) | b \rangle
\Big]   \nonumber \\
&
+
\Delta E_{\rm 1ph}
\Big[
\langle a | {\rm P}'(\veps_a) | a \rangle
+
\langle b | {\rm P}'(\veps_b) | b \rangle
\Big] \, ,
\end{align}
where $I'(\Delta)=dI/d\omega |_{\omega=\Delta}$, ${\rm P}'(\veps_a) = d{\rm P}(E)/dE |_{E=\veps_a}$, and $\Delta E_{\rm 1ph} = \sum_P (-1)^P \matr{PaPb}{I(\Delta)}{ab}$ is the one-photon-exchange correction. Finally, the vertex contribution is given by
\begin{widetext}
\begin{align}
\label{eq:vert_1el}
\Delta E_{\rm vert}^{(2)} &=
\sum_P (-1)^P  
\frac{i}{2\pi} \int\limits_{-\infty}^{\infty} \! d\omega \, \sum_{n_1n_2}
\Bigg[
\frac{ \matr{Pa\, n_2}{R(\omega)}{n_1a} \matr{n_1 Pb}{I(\Delta)}{n_2b} }
     { (\omega+\veps_{Pa}-u\veps_{n_1})(\omega+\veps_a-u\veps_{n_2}) }   
+ 
\frac{ \matr{Pa\, n_1}{I(\Delta)}{an_2} \matr{Pb\, n_2}{R(\omega)}{n_1b} }
     { (\omega+\veps_{Pb}-u\veps_{n_1})(\omega+\veps_b-u\veps_{n_2}) } 
\Bigg ]    \, . 
\end{align}
To summarize, within the rigorous QED approach the interelectronic-interaction correction of first order in $1/Z$ to the one-electron part of the nuclear recoil effect is given by the sum of Eqs.~(\ref{eq:irr_1el}), (\ref{eq:red_1el}), and (\ref{eq:vert_1el}). The calculations are to be performed for all the operators~(\ref{eq:R_c})--(\ref{eq:R_tr2}),
\begin{align}
\label{eq:dE_rec2}
\Delta E^{(2)}_{\rm 1el} = \Delta E^{(2)}_{\rm c} + \Delta E^{(2)}_{\rm tr1} + \Delta E^{(2)}_{\rm tr2} \, .
\end{align}

As in the case of the independent-electron approximation discussed in Sec.~\ref{sec:1}, the integration over $\omega$ in the Coulomb contribution~$\Delta E^{(2)}_{\rm c}$ can be carried out analytically. The irreducible contribution and the part of the reducible contribution with $I'$ can be treated similar to Eq.~(\ref{eq:dE_c_1el}) using the formula (\ref{eq:sokhot}). These terms can be rewritten in the form similar to Eq.~(\ref{eq:dE_c_1el_calc}). For the other contributions, Cauchy's residue theorem should be employed. Then, the second part of the reducible contribution (with the operator ${\rm P'}$) vanishes, since it contains only the second-order poles for the Coulomb interaction (\ref{eq:R_c}). Finally, the vertex contribution is
\begin{align}
\label{eq:vert_c_1el}
\Delta E_{\rm c,vert}^{(2)} &= \frac{1}{M} \sum_P (-1)^P 
\Bigg\{
\sum_{n_1}^{\veps_{n_1}<0} \! \sum^{n_2}_{\veps_{n_2}>0} 
\Bigg[
\frac{ \matr{Pa}{p_k}{n_1} \matr{n_2}{p_k}{a} \matr{n_1 Pb}{I(\Delta)}{n_2b} }
     { \veps_{n_2} - \veps_{n_1} + \Delta }
+
\frac{ \matr{Pa\, n_1}{I(\Delta)}{an_2} \matr{Pb}{p_k}{n_1} \matr{n_2}{p_k}{b} }
     { \veps_{n_2} - \veps_{n_1} - \Delta }
\Bigg]   \nonumber \\
& 
+
\sum_{n_2}^{\veps_{n_2}<0} \! \sum^{n_1}_{\veps_{n_1}>0} 
\Bigg[
\frac{ \matr{Pa}{p_k}{n_1} \matr{n_2}{p_k}{a} \matr{n_1 Pb}{I(\Delta)}{n_2b} }
     { \veps_{n_1} - \veps_{n_2} - \Delta }
+
\frac{ \matr{Pa\, n_1}{I(\Delta)}{an_2} \matr{Pb}{p_k}{n_1} \matr{n_2}{p_k}{b} }
     { \veps_{n_1} - \veps_{n_2} + \Delta }
\Bigg] 
\Bigg\} \, .
\end{align}
\end{widetext}

Concluding this section, we note that the Breit-approximation results for the electron-electron correction to the NMS can be obtained from the QED formulas derived in the present work. To do so, one has to neglect the energy dependence in the operators $\bD(\omega)$ and $I(\omega)$ in Eqs.~(\ref{eq:D}) and (\ref{eq:I_coul}), respectively, and introduce the projectors on the positive-energy part of the spectrum in Eqs.~(\ref{eq:xi_a}), (\ref{eq:xi_b}), and (\ref{eq:vert_1el}). In addition, the lowest-order relativistic limit of the operator ${\rm P}$ in Eq.~(\ref{eq:P_operator}) has to be treated as discussed in Sec.~\ref{sec:1}. On these assumptions, the integration over $\omega$ in all the expressions can be performed analytically by employing Eq.~(\ref{eq:sokhot}) and Cauchy's residue theorem. As a result, the reducible and vertex contributions vanish identically, and the irreducible contribution reproduces the interelectronic-interaction correction of first order in $1/Z$ to the one-electron part of the nuclear recoil within the Breit approximation. Obviously, the two-transverse-photon contribution has to be omitted in this approximation.


\section{Numerical results and discussion \label{sec:3}}

\begin{table*}[t]
\centering

\renewcommand{\arraystretch}{1.1}

\caption{\label{tab:0:all_1el_contrib1s_2s_2p1} 
         The nuclear recoil contribution to binding energies of the $1s$, $2s$, and $2p_{1/2}$ states 
         expressed in terms of the function $A(\alpha Z)$ defined by Eq.~(\ref{eq:A_alphaZ}).
         For each $Z$, the first line shows the results of the QED calculations to all orders in $\alpha Z$, 
         whereas the second line displays the values obtained within the Breit approximation employing 
         the normal mass shift (NMS) operator given in Eq.~(\ref{eq:NMS}). The individual contributions, 
         the Coulomb ($\rm c$), the one-transverse-photon ($\rm tr1$), and the two-transverse-photon ($\rm tr2$) ones, 
         are shown only for the $1s$ state. 
         }
         
\begin{tabular}{@{}
                l
                c
                S[table-format= 4.7]
                S[table-format=-1.7]
                S[table-format=-1.7]
                S[table-format= 1.8]
                S[table-format= 3.8]
                S[table-format= 3.6]@{}
               }
               
\hline
\hline 
 
   \multicolumn{1}{c}{\rule{0pt}{1.2em}$Z$}              &
   \multicolumn{1}{c}{Approach}                            &
   \multicolumn{1}{c}{~$A_{\rm c}^{1s}(\alpha Z)$}        &
   \multicolumn{1}{c}{$A_{\rm tr1}^{1s}(\alpha Z)$}       &
   \multicolumn{1}{c}{$A_{\rm tr2}^{1s}(\alpha Z)$}       &
   \multicolumn{1}{c}{$A^{1s}(\alpha Z)$\,\,\,\,}         &
   \multicolumn{1}{c}{$A^{2s}(\alpha Z)$}                 &
   \multicolumn{1}{c}{\,\,\,$A^{2p_{1/2}}(\alpha Z)$}     \\
        
\hline   
                       
  \multirow{2}{*}{  5}  &  QED \rule{0pt}{2.6ex} &       0.501315   &      -0.001201   &      -0.000047   &       0.500066   &       0.125051   &       0.125041   \\ 
                        &  $H_{\rm NMS}$  &       0.501333   &      -0.001334   &   {\text{---}}   &       0.499999   &       0.125042   &       0.125042   \\[1.5mm] 

  \multirow{2}{*}{ 10}  &  QED            &       0.505223   &      -0.004502   &      -0.000253   &       0.500468   &       0.125237   &       0.125163   \\ 
                        &  $H_{\rm NMS}$  &       0.505363   &      -0.005368   &   {\text{---}}   &       0.499995   &       0.125166   &       0.125167   \\[1.5mm] 

  \multirow{2}{*}{ 20}  &  QED            &       0.520945   &      -0.016490   &      -0.001206   &       0.503248   &       0.126176   &       0.125656   \\ 
                        &  $H_{\rm NMS}$  &       0.521953   &      -0.021999   &   {\text{---}}   &       0.499954   &       0.125667   &       0.125673   \\[1.5mm] 

  \multirow{2}{*}{ 30}  &  QED            &       0.548214   &      -0.035322   &      -0.002787   &       0.510105   &       0.128157   &       0.126517   \\ 
                        &  $H_{\rm NMS}$  &       0.551392   &      -0.051594   &   {\text{---}}   &       0.499798   &       0.126508   &       0.126534   \\[1.5mm] 

  \multirow{2}{*}{ 40}  &  QED            &       0.589533   &      -0.061804   &      -0.004789   &       0.522940   &       0.131600   &       0.127850   \\ 
                        &  $H_{\rm NMS}$  &       0.596793   &      -0.097421   &   {\text{---}}   &       0.499373   &       0.127698   &       0.127781   \\[1.5mm] 

  \multirow{2}{*}{ 50}  &  QED            &       0.649386   &      -0.098199   &      -0.006909   &       0.544277   &       0.137127   &       0.129846   \\ 
                        &  $H_{\rm NMS}$  &       0.663456   &      -0.165147   &   {\text{---}}   &       0.498309   &       0.129224   &       0.129456   \\[1.5mm] 

  \multirow{2}{*}{ 60}  &  QED            &       0.735510   &      -0.148884   &      -0.008655   &       0.577971   &       0.145734   &       0.132852   \\ 
                        &  $H_{\rm NMS}$  &       0.760401   &      -0.264500   &   {\text{---}}   &       0.495901   &       0.131033   &       0.131619   \\[1.5mm] 

  \multirow{2}{*}{ 70}  &  QED            &       0.860884   &      -0.221722   &      -0.009082   &       0.630080   &       0.159089   &       0.137493   \\ 
                        &  $H_{\rm NMS}$  &       0.902512   &      -0.412217   &   {\text{---}}   &       0.490294   &       0.132886   &       0.134339   \\[1.5mm] 

  \multirow{2}{*}{ 80}  &  QED            &       1.049608   &      -0.331694   &      -0.006266   &       0.711648   &       0.180304   &       0.144991   \\ 
                        &  $H_{\rm NMS}$  &       1.117655   &      -0.639407   &   {\text{---}}   &       0.478248   &       0.134239   &       0.137667   \\[1.5mm] 

  \multirow{2}{*}{ 90}  &  QED            &       1.343409   &      -0.506445   &       0.004025   &       0.840989   &       0.215024   &       0.157885   \\ 
                        &  $H_{\rm NMS}$  &       1.453046   &      -1.002570   &   {\text{---}}   &       0.450476   &       0.133231   &       0.141468   \\[1.5mm] 

  \multirow{2}{*}{ 92}  &  QED            &       1.420806   &      -0.553271   &       0.007663   &       0.875198   &       0.224427   &       0.161526   \\ 
                        &  $H_{\rm NMS}$  &       1.541376   &      -1.099804   &   {\text{---}}   &       0.441572   &       0.132411   &       0.142234   \\[1.5mm] 

  \multirow{2}{*}{ 95}  &  QED            &       1.553578   &      -0.634200   &       0.014619   &       0.933997   &       0.240743   &       0.167966   \\ 
                        &  $H_{\rm NMS}$  &       1.693024   &      -1.267592   &   {\text{---}}   &       0.425432   &       0.130603   &       0.143340   \\[1.5mm] 

  \multirow{2}{*}{100}  &  QED            &       1.831537   &      -0.805924   &       0.031669   &       1.057282   &       0.275548   &       0.182201   \\ 
                        &  $H_{\rm NMS}$  &       2.010789   &      -1.622347   &   {\text{---}}   &       0.388442   &       0.125349   &       0.144890   \\[0.5mm]

\hline
\hline

\end{tabular}%

\end{table*}

In this work, the formalism derived in Sec.~\ref{sec:1} and \ref{sec:2} is employed for nonperturbative (in $\alpha Z$) calculations of the one-electron contribution to the nuclear recoil effect on binding energies of the $1s^2$ state in heliumlike ions and the $1s^2 2s$ and $1s^2 2p_{1/2}$ states in lithiumlike ions. By evaluating the differences of the results obtained for the binding energies, one can calculate the corresponding contributions to the ionization energies of the $1s^2 2s$ and $1s^2 2p_{1/2}$ states and to the $2p_{1/2}$--$2s$ transition energy in Li-like ions. The calculations are performed in the range $Z=5$--$100$. As noted in Sec.~\ref{sec:2}, for the Coulomb contributions the integration over $\omega$ can be performed analytically. For the one-transverse-photon and two-transverse-photon contributions, the corresponding integrals are calculated numerically employing Wick's rotation of the integration contour to the complex plane; see Ref.~\cite{Malyshev:2020:012513} for the details. The summation over the one-electron states in the electron propagator is carried out using the finite basis set of the Dirac-equation eigenfunctions constructed from the B-splines~\cite{Johnson:1988:307, Sapirstein:1996:5213} by means of the dual-kinetic-balance approach~\cite{splines:DKB}. 

Within the Furry picture, the finite nuclear size correction to various atomic properties generally can be taken into account by substituting the potential of the extended nucleus into the Dirac equation~(\ref{eq:DirEq}). In the case of the nuclear recoil effect, this recipe leads only to a partial treatment of the nuclear size correction~\cite{Shabaev:1998:59}. The rigorous evaluation of this correction has been performed up to date only within the Breit approximation~\cite{Grotch:1969:350, Borie:1982:67, Aleksandrov:2015:144004}. The discussion of the uncertainty related to this approximate treatment of the nuclear size correction to the nuclear recoil effect can be found, e.g., in Refs.~\cite{Yerokhin:2015:033103, Malyshev:2018:085001}. We stress that this uncertainty exceeds the one which can be obtained by varying the nuclear charge distribution model and the nuclear charge radius within its error bar. In the present study, the Fermi model for the nuclear charge distribution is used for all nuclei with $Z\geqslant 15$. Otherwise, the homogeneously charged-sphere model is employed. The nuclear charge radii are taken from Refs.~\cite{Angeli:2013:69, Yerokhin:2015:033103}.

\subsection{One-electron part of the nuclear recoil effect}

\begin{table}[t]
\centering

\renewcommand{\arraystretch}{1.1}

\caption{\label{tab:1:bind_1s1s} 
         The interelectronic-interaction correction of first order in $1/Z$ to the one-electron part
         of the nuclear recoil contribution to the binding energy of the $1s^2$ state
         expressed in terms of the dimensionless function $B(\alpha Z)$ defined by Eq.~(\ref{eq:B_alphaZ}).  
         }
         
\resizebox{\columnwidth}{!}{%
\begin{tabular}{@{}
                l
                c
                S[table-format=-1.5(1)]
                S[table-format= 1.5(1)]
                S[table-format=-1.5(1)]
                S[table-format=-1.5(1)]@{}
               }
               
\hline
\hline

   \multicolumn{1}{c}{\rule{0pt}{1.2em}$Z$}      &
   \multicolumn{1}{c}{Approach}                   &
   \multicolumn{1}{c}{$B_{\rm c  }(\alpha Z)$}   &
   \multicolumn{1}{c}{$B_{\rm tr1}(\alpha Z)$}   &
   \multicolumn{1}{c}{$B_{\rm tr2}(\alpha Z)$}   &
   \multicolumn{1}{c}{$B(\alpha Z)$}             \\      
        
\hline   
                       
  \multirow{2}{*}{  5}  &  QED \rule{0pt}{2.6ex} &       -0.62826   &        0.00185   &        0.00006   &       -0.62635   \\ 
                        &  $H_{\rm NMS}$  &       -0.62829   &        0.00207   &   {\text{---}}   &       -0.62622   \\[1.5mm] 

  \multirow{2}{*}{ 10}  &  QED            &       -0.63805   &        0.00695   &        0.00031   &       -0.63080   \\ 
                        &  $H_{\rm NMS}$  &       -0.63824   &        0.00837   &   {\text{---}}   &       -0.62987   \\[1.5mm] 

  \multirow{2}{*}{ 20}  &  QED            &       -0.67814   &        0.02589   &        0.00141   &       -0.65084   \\ 
                        &  $H_{\rm NMS}$  &       -0.67952   &        0.03512   &   {\text{---}}   &       -0.64440   \\[1.5mm] 

  \multirow{2}{*}{ 30}  &  QED            &       -0.74943   &        0.05739   &        0.00312   &       -0.68892   \\ 
                        &  $H_{\rm NMS}$  &       -0.75394   &        0.08562   &   {\text{---}}   &       -0.66832   \\[1.5mm] 

  \multirow{2}{*}{ 40}  &  QED            &       -0.86063   &        0.10551   &        0.00508   &       -0.75005   \\ 
                        &  $H_{\rm NMS}$  &       -0.87142   &        0.17013   &   {\text{---}}   &       -0.70129   \\[1.5mm] 

  \multirow{2}{*}{ 50}  &  QED            &       -1.02728   &        0.17855   &        0.00662   &       -0.84211   \\ 
                        &  $H_{\rm NMS}$  &       -1.04937   &        0.30672   &   {\text{---}}   &       -0.74264   \\[1.5mm] 

  \multirow{2}{*}{ 60}  &  QED            &       -1.27663   &     0.29171(1)   &        0.00643   &       -0.97848   \\ 
                        &  $H_{\rm NMS}$  &       -1.31831   &        0.52716   &   {\text{---}}   &       -0.79115   \\[1.5mm] 

  \multirow{2}{*}{ 70}  &  QED            &       -1.65677   &     0.47294(1)   &        0.00162   &    -1.18220(1)   \\ 
                        &  $H_{\rm NMS}$  &       -1.73185   &        0.88874   &   {\text{---}}   &       -0.84311   \\[1.5mm] 

  \multirow{2}{*}{ 80}  &  QED            &       -2.25999   &     0.77744(1)   &       -0.01449   &       -1.49704   \\ 
                        &  $H_{\rm NMS}$  &       -2.39366   &        1.50299   &   {\text{---}}   &       -0.89067   \\[1.5mm] 

  \multirow{2}{*}{ 90}  &  QED            &       -3.26153   &     1.31583(1)   &       -0.05794   &    -2.00364(1)   \\ 
                        &  $H_{\rm NMS}$  &       -3.49868   &        2.59077   &   {\text{---}}   &       -0.90791   \\[1.5mm] 

  \multirow{2}{*}{ 92}  &  QED            &       -3.53632   &     1.46896(1)   &       -0.07270   &    -2.14006(1)   \\ 
                        &  $H_{\rm NMS}$  &       -3.80258   &        2.89955   &   {\text{---}}   &       -0.90304   \\[1.5mm] 

  \multirow{2}{*}{ 95}  &  QED            &       -4.01641   &     1.74072(1)   &       -0.10079   &    -2.37648(1)   \\ 
                        &  $H_{\rm NMS}$  &       -4.33457   &        3.44668   &   {\text{---}}   &       -0.88788   \\[1.5mm] 

  \multirow{2}{*}{100}  &  QED            &    -5.05385(1)   &     2.34318(4)   &    -0.16993(1)   &    -2.88060(3)   \\ 
                        &  $H_{\rm NMS}$  &       -5.48748   &        4.65565   &   {\text{---}}   &       -0.83183   \\[0.5mm]

\hline
\hline

\end{tabular}%
}

\end{table}

\begin{table}[t]
\centering

\renewcommand{\arraystretch}{1.1}

\caption{\label{tab:1:bind_1s1s2s} 
         The interelectronic-interaction correction of first order in $1/Z$ to the one-electron part
         of the nuclear recoil contribution to the binding energy of the $1s^22s$ state
         expressed in terms of the dimensionless function $B(\alpha Z)$ defined by Eq.~(\ref{eq:B_alphaZ}).  
         }
         
\resizebox{\columnwidth}{!}{%
\begin{tabular}{@{}
                l
                c
                S[table-format=-1.5(1)]
                S[table-format= 1.5(1)]
                S[table-format=-1.5(1)]
                S[table-format=-1.5(1)]@{}
               }
               
\hline
\hline

   \multicolumn{1}{c}{\rule{0pt}{1.2em}$Z$}      &
   \multicolumn{1}{c}{Approach}                   &
   \multicolumn{1}{c}{$B_{\rm c  }(\alpha Z)$}   &
   \multicolumn{1}{c}{$B_{\rm tr1}(\alpha Z)$}   &
   \multicolumn{1}{c}{$B_{\rm tr2}(\alpha Z)$}   &
   \multicolumn{1}{c}{$B(\alpha Z)$}             \\      
        
\hline   
                       
  \multirow{2}{*}{  5}  &  QED \rule{0pt}{2.6ex} &       -1.02742   &        0.00274   &        0.00008   &       -1.02460   \\ 
                        &  $H_{\rm NMS}$  &       -1.02745   &        0.00303   &   {\text{---}}   &       -1.02442   \\[1.5mm] 

  \multirow{2}{*}{ 10}  &  QED            &       -1.04126   &        0.01036   &        0.00041   &       -1.03049   \\ 
                        &  $H_{\rm NMS}$  &       -1.04151   &        0.01224   &   {\text{---}}   &       -1.02927   \\[1.5mm] 

  \multirow{2}{*}{ 20}  &  QED            &       -1.09798   &        0.03896   &        0.00185   &       -1.05716   \\ 
                        &  $H_{\rm NMS}$  &       -1.09983   &        0.05122   &   {\text{---}}   &       -1.04861   \\[1.5mm] 

  \multirow{2}{*}{ 30}  &  QED            &       -1.19891   &        0.08672   &        0.00399   &       -1.10820   \\ 
                        &  $H_{\rm NMS}$  &       -1.20494   &        0.12427   &   {\text{---}}   &       -1.08067   \\[1.5mm] 

  \multirow{2}{*}{ 40}  &  QED            &       -1.35644   &        0.15948   &        0.00614   &       -1.19082   \\ 
                        &  $H_{\rm NMS}$  &       -1.37084   &        0.24554   &   {\text{---}}   &       -1.12530   \\[1.5mm] 

  \multirow{2}{*}{ 50}  &  QED            &       -1.59271   &        0.26916   &        0.00717   &       -1.31638   \\ 
                        &  $H_{\rm NMS}$  &       -1.62219   &        0.44013   &   {\text{---}}   &       -1.18206   \\[1.5mm] 

  \multirow{2}{*}{ 60}  &  QED            &       -1.94675   &     0.43776(1)   &        0.00484   &       -1.50415   \\ 
                        &  $H_{\rm NMS}$  &       -2.00248   &        0.75257   &   {\text{---}}   &       -1.24992   \\[1.5mm] 

  \multirow{2}{*}{ 70}  &  QED            &       -2.48789   &     0.70598(1)   &       -0.00556   &    -1.78748(1)   \\ 
                        &  $H_{\rm NMS}$  &       -2.58861   &        1.26394   &   {\text{---}}   &       -1.32467   \\[1.5mm] 

  \multirow{2}{*}{ 80}  &  QED            &       -3.35000   &     1.15466(1)   &       -0.03461   &       -2.22995   \\ 
                        &  $H_{\rm NMS}$  &       -3.53029   &        2.13366   &   {\text{---}}   &       -1.39663   \\[1.5mm] 

  \multirow{2}{*}{ 90}  &  QED            &       -4.79042   &     1.94762(1)   &       -0.10743   &    -2.95022(1)   \\ 
                        &  $H_{\rm NMS}$  &       -5.11280   &        3.68168   &   {\text{---}}   &       -1.43112   \\[1.5mm] 

  \multirow{2}{*}{ 92}  &  QED            &       -5.18740   &     2.17346(1)   &       -0.13159   &    -3.14553(1)   \\ 
                        &  $H_{\rm NMS}$  &       -5.55004   &        4.12298   &   {\text{---}}   &       -1.42707   \\[1.5mm] 

  \multirow{2}{*}{ 95}  &  QED            &       -5.88241   &     2.57461(2)   &       -0.17719   &    -3.48500(1)   \\ 
                        &  $H_{\rm NMS}$  &       -6.31700   &        4.90663   &   {\text{---}}   &       -1.41037   \\[1.5mm] 

  \multirow{2}{*}{100}  &  QED            &    -7.38956(1)   &     3.46536(4)   &    -0.28818(1)   &    -4.21237(3)   \\ 
                        &  $H_{\rm NMS}$  &       -7.98510   &        6.64491   &   {\text{---}}   &       -1.34018   \\[0.5mm]

\hline
\hline

\end{tabular}%
}

\end{table}

\begin{table}[t]
\centering

\renewcommand{\arraystretch}{1.1}

\caption{\label{tab:1:bind_1s1s2p1} 
         The interelectronic-interaction correction of first order in $1/Z$ to the one-electron part
         of the nuclear recoil contribution to the binding energy of the $1s^22p_{1/2}$ state
         expressed in terms of the dimensionless function $B(\alpha Z)$ defined by Eq.~(\ref{eq:B_alphaZ}).  
         }
         
\resizebox{\columnwidth}{!}{%
\begin{tabular}{@{}
                l
                c
                S[table-format=-1.5(1)]
                S[table-format= 1.5(1)]
                S[table-format=-1.5(1)]
                S[table-format=-1.5(1)]@{}
               }
               
\hline
\hline

   \multicolumn{1}{c}{\rule{0pt}{1.2em}$Z$}      &
   \multicolumn{1}{c}{Approach}                   &
   \multicolumn{1}{c}{$B_{\rm c  }(\alpha Z)$}   &
   \multicolumn{1}{c}{$B_{\rm tr1}(\alpha Z)$}   &
   \multicolumn{1}{c}{$B_{\rm tr2}(\alpha Z)$}   &
   \multicolumn{1}{c}{$B(\alpha Z)$}             \\      
        
\hline   
                       
  \multirow{2}{*}{  5}  &  QED \rule{0pt}{2.6ex} &       -1.09828   &        0.00266   &        0.00007   &       -1.09556   \\ 
                        &  $H_{\rm NMS}$  &       -1.09831   &        0.00289   &   {\text{---}}   &       -1.09542   \\[1.5mm] 

  \multirow{2}{*}{ 10}  &  QED            &       -1.11258   &        0.01017   &        0.00034   &       -1.10207   \\ 
                        &  $H_{\rm NMS}$  &       -1.11278   &        0.01169   &   {\text{---}}   &       -1.10109   \\[1.5mm] 

  \multirow{2}{*}{ 20}  &  QED            &       -1.17123   &        0.03906   &        0.00150   &       -1.13066   \\ 
                        &  $H_{\rm NMS}$  &       -1.17275   &        0.04899   &   {\text{---}}   &       -1.12375   \\[1.5mm] 

  \multirow{2}{*}{ 30}  &  QED            &       -1.27558   &        0.08863   &        0.00305   &       -1.18391   \\ 
                        &  $H_{\rm NMS}$  &       -1.28056   &        0.11905   &   {\text{---}}   &       -1.16151   \\[1.5mm] 

  \multirow{2}{*}{ 40}  &  QED            &       -1.43815   &        0.16579   &        0.00404   &       -1.26832   \\ 
                        &  $H_{\rm NMS}$  &       -1.45015   &        0.23572   &   {\text{---}}   &       -1.21443   \\[1.5mm] 

  \multirow{2}{*}{ 50}  &  QED            &       -1.68116   &        0.28397   &        0.00274   &       -1.39445   \\ 
                        &  $H_{\rm NMS}$  &       -1.70605   &        0.42351   &   {\text{---}}   &       -1.28254   \\[1.5mm] 

  \multirow{2}{*}{ 60}  &  QED            &       -2.04377   &     0.46749(1)   &       -0.00418   &       -1.58047   \\ 
                        &  $H_{\rm NMS}$  &       -2.09154   &        0.72595   &   {\text{---}}   &       -1.36559   \\[1.5mm] 

  \multirow{2}{*}{ 70}  &  QED            &       -2.59593   &     0.76112(1)   &       -0.02356   &    -1.85838(1)   \\ 
                        &  $H_{\rm NMS}$  &       -2.68384   &        1.22278   &   {\text{---}}   &       -1.46106   \\[1.5mm] 

  \multirow{2}{*}{ 80}  &  QED            &       -3.47332   &     1.25346(1)   &       -0.07027   &       -2.29013   \\ 
                        &  $H_{\rm NMS}$  &       -3.63420   &        2.07194   &   {\text{---}}   &       -1.56226   \\[1.5mm] 

  \multirow{2}{*}{ 90}  &  QED            &       -4.94405   &     2.12708(1)   &       -0.17879   &    -2.99576(1)   \\ 
                        &  $H_{\rm NMS}$  &       -5.24001   &        3.59851   &   {\text{---}}   &       -1.64151   \\[1.5mm] 

  \multirow{2}{*}{ 92}  &  QED            &       -5.35164   &     2.37701(1)   &       -0.21386   &    -3.18849(1)   \\ 
                        &  $H_{\rm NMS}$  &       -5.68680   &        4.03754   &   {\text{---}}   &       -1.64926   \\[1.5mm] 

  \multirow{2}{*}{ 95}  &  QED            &       -6.06731   &     2.82181(1)   &       -0.27938   &    -3.52488(1)   \\ 
                        &  $H_{\rm NMS}$  &       -6.47332   &        4.82083   &   {\text{---}}   &       -1.65249   \\[1.5mm] 

  \multirow{2}{*}{100}  &  QED            &    -7.62960(1)   &     3.81396(4)   &    -0.43661(1)   &    -4.25225(3)   \\ 
                        &  $H_{\rm NMS}$  &       -8.19725   &        6.57466   &   {\text{---}}   &       -1.62259   \\[0.5mm]

\hline
\hline

\end{tabular}%
}

\end{table}

We start with the results obtained within the independent-electron approximation. As noted above, numerous calculations of the one-electron contribution to the nuclear recoil effect on binding energies of low-lying states in hydrogenlike ions can be found in the literature; see  Refs.~\cite{Artemyev:1995:1884, Artemyev:1995:5201, Adkins:2007:042508, Malyshev:2018:085001}. Nevertheless, for the sake of completeness, we summarize our results for the one-electron contribution to zeroth order in $1/Z$ in Table~\ref{tab:0:all_1el_contrib1s_2s_2p1}. The results for the $1s$, $2s$, and $2p_{1/2}$ states are given in terms of the dimensionless function $A(\alpha Z)$ defined by
\begin{equation}
\label{eq:A_alphaZ}
\Delta E^{(1)} = \frac{m}{M}(\alpha Z)^2 A(\alpha Z) \, mc^2 \, .
\end{equation}
To avoid misunderstanding, we note that the index ``(1)'' here (and analogous indices below) refers to the perturbation-theory order in the framework of the TTGF method, and it is equal to the order in $1/Z$ plus one. For each $Z$ (in Table~\ref{tab:0:all_1el_contrib1s_2s_2p1}), the values calculated within the rigorous QED formalism employing Eqs.~(\ref{eq:dE_c_1el})--(\ref{eq:dE_tr2_1el}) are displayed in the first line, while the results obtained by means of the NMS operator (\ref{eq:NMS}) are shown in the second line. The Coulomb~$A_{\rm c}$, the one-transverse-photon~$A_{\rm tr1}$, and the two-transverse-photon~$A_{\rm tr2}$ contributions are shown only for the $1s$ state in order to provide insight into how the individual terms contribute to the total values. From Table~\ref{tab:0:all_1el_contrib1s_2s_2p1}, one can see that the nontrivial QED part of the nuclear recoil effect can significantly alter the Breit-approximation result. For instance, for the $1s$ state it even exceeds the lowest-order relativistic value for $Z>92$.

The first-order (in $1/Z$) electron-electron interaction correction to the nuclear recoil effect can be conveniently expressed in terms of the dimensionless function $B(\alpha Z)$ defined by
\begin{equation}
\label{eq:B_alphaZ}
\Delta E^{(2)} = \frac{m}{M}\frac{(\alpha Z)^2}{Z} B(\alpha Z) \, mc^2 \, .
\end{equation}
Our results for the interelectronic-interaction correction to the one-electron part of the nuclear recoil effect on the binding energies of the $1s^2$, $1s^2 2s$, and $1s^2 2p_{1/2}$ states are shown in Tables~\ref{tab:1:bind_1s1s}, \ref{tab:1:bind_1s1s2s}, and \ref{tab:1:bind_1s1s2p1}, respectively. As in Table~\ref{tab:0:all_1el_contrib1s_2s_2p1}, for each $Z$ we present two values. The first value is evaluated within the framework of the \textit{ab initio} approach derived in the preceding section, whereas the second one is obtained within the Breit approximation via the NMS operator (\ref{eq:NMS}). The functions $B_{\rm c}$, $B_{\rm tr1}$, and $B_{\rm tr2}$ correspond to the contributions of the Coulomb~(\ref{eq:R_c}), the one-transverse-photon~(\ref{eq:R_tr1}), and the two-transverse-photon~(\ref{eq:R_tr2}) interactions, respectively. The uncertainties given in the parentheses are due to the numerical errors only. They are estimated by studying the convergence of the results with respect to the size of the basis set as well as the number of points in the quadrature formula for the integration over $\omega$.

\begin{figure}
\begin{center}
\includegraphics[width=\columnwidth]{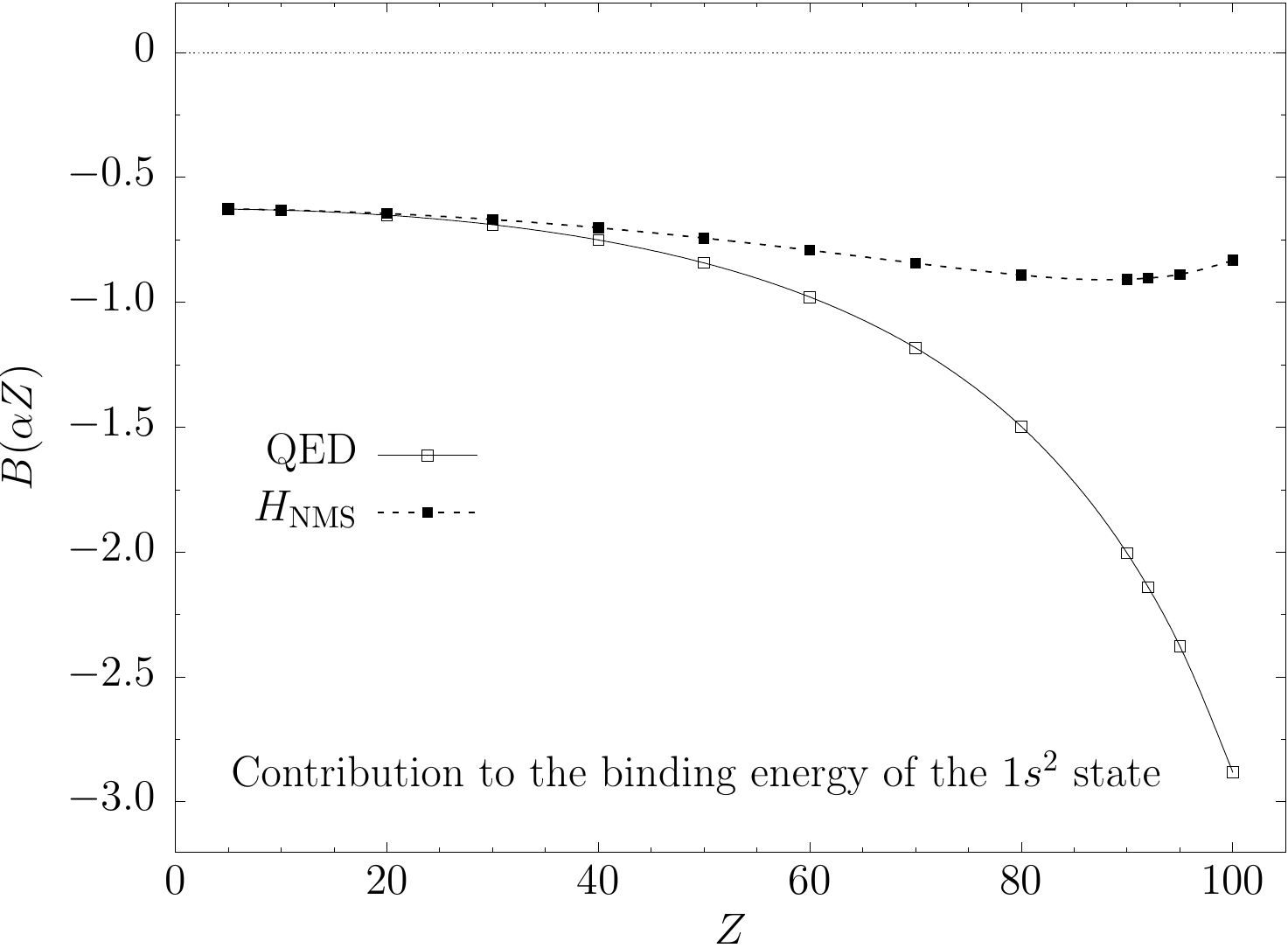}
\caption{\label{fig:1:bind_1s1s}
The interelectronic-interaction correction of first order in $1/Z$ to the one-electron part of the nuclear recoil effect on the binding energy of the $1s^2$ state expressed in terms of the dimensionless function $B(\alpha Z)$ defined by Eq.~(\ref{eq:B_alphaZ}). The solid line represents the results of the QED calculations to all orders in $\alpha Z$, whereas the dashed line corresponds to the calculations based on the normal mass shift (NMS) operator given by Eq.~(\ref{eq:NMS}).}
\end{center}
\end{figure}

\begin{figure}
\begin{center}
\includegraphics[width=\columnwidth]{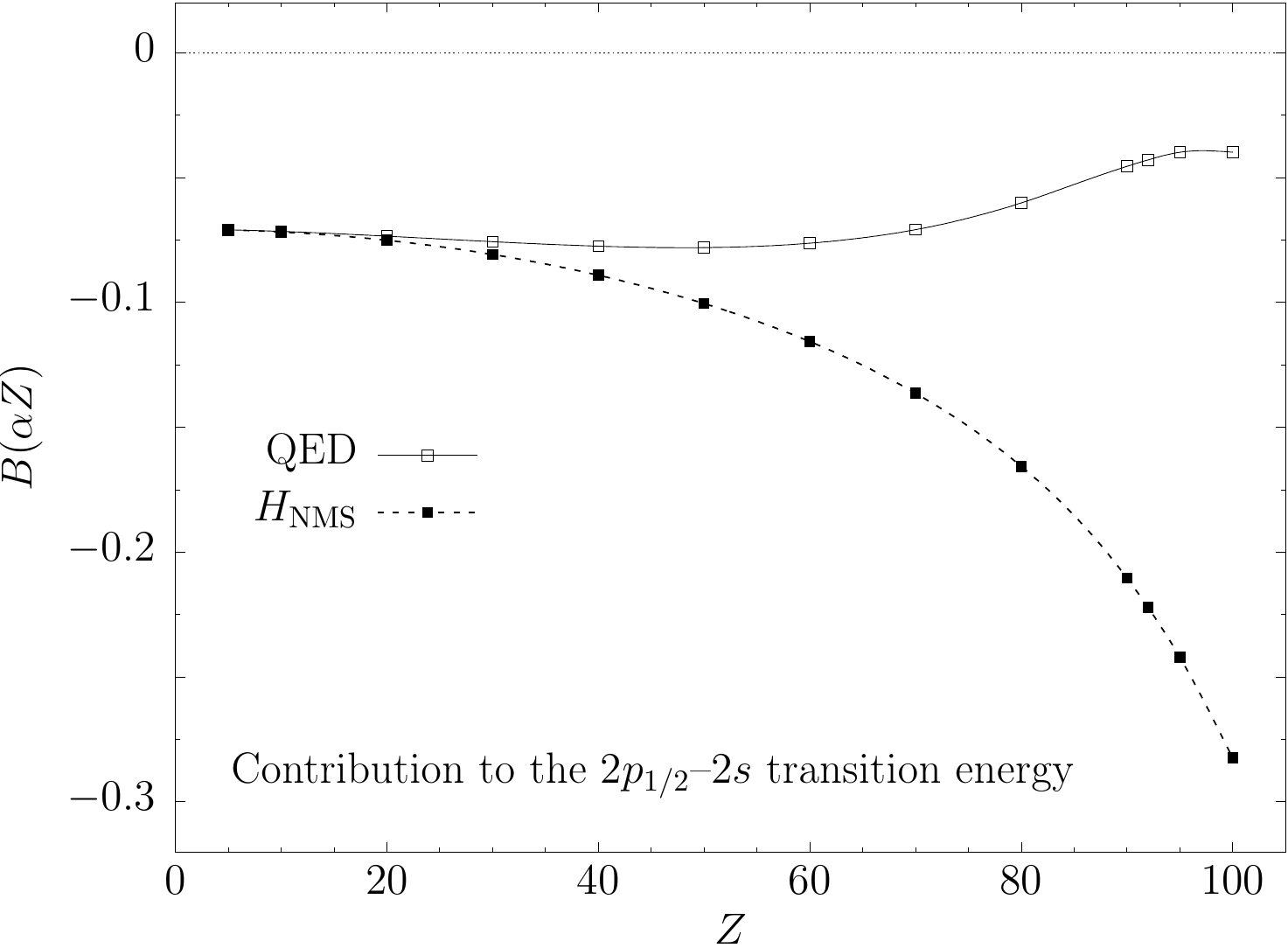}
\caption{\label{fig:1:tr_2p1_2s}
The interelectronic-interaction correction of first order in $1/Z$ to the one-electron part of the nuclear recoil effect on the $2p_{1/2}$--$2s$ transition energy in Li-like ions expressed in terms of the dimensionless function $B(\alpha Z)$ defined by Eq.~(\ref{eq:B_alphaZ}). Notations are the same as in Fig.~\ref{fig:1:bind_1s1s}.}
\end{center}
\end{figure}

From Tables~\ref{tab:1:bind_1s1s}--\ref{tab:1:bind_1s1s2p1}, one can see that the results of the QED calculations tend to the Breit-approximation values when $\alpha Z\rightarrow 0$. This behavior is what one can expect, having in mind that the NMS operator (\ref{eq:NMS}) provides the lowest-order relativistic approximation to the theory worked out. On the other hand, due to the energy dependence of the vector $\bD(\omega)$ in the integration over $\omega$ in Eq.~(\ref{eq:P_operator}) and analogous expressions, the one-transverse-photon contribution acquires the considerable correction compared to the Breit approximation for high-$Z$ ions. The nontrivial QED Coulomb contribution as well as the two-transverse-photon contribution also grow rapidly with increasing $Z$. As a result, the higher orders (in $\alpha Z$) modify the behavior of the function $B(\alpha Z)$ significantly. Indeed, the function $B(\alpha Z)$ calculated to all orders in $\alpha Z$ may differ by several times from the approximate one obtained by means of the NMS operator. In order to illustrate this fact, the interelectronic-interaction correction to the one-electron part of the nuclear recoil effect on the binding energy of the $1s^2$ state is plotted in Fig.~\ref{fig:1:bind_1s1s}, where the data given in the last column of Table~\ref{tab:1:bind_1s1s} are presented. The Breit-approximation values and the \textit{ab initio} QED results are shown with the dashed and solid lines, respectively. It is seen that the NMS operator leads to the strong underestimation of the nuclear recoil effect at the high-$Z$ region. The similar situation takes place for binding energies of the $1s^2 2s$ and $1s^2 2p_{1/2}$ states. However, it is not always the case. For instance, the contribution under consideration to the $2p_{1/2}$--$2s$ transition energy in Li-like ions is presented in Fig.~\ref{fig:1:tr_2p1_2s}. In this transition energy, the interelectronic-interaction correction to the one-electron part of the nuclear recoil effect obtained by means of the rigorous QED approach appears to be less pronounced than the one evaluated within the lowest-order relativistic approximation.

\subsection{Total nuclear recoil effect to first order in $1/Z$}

\begin{table*}[t]
\centering

\renewcommand{\arraystretch}{1.1}

\caption{\label{tab:01:bind_1s1s} 
         The nuclear recoil contribution to the binding energy of the $1s^2$ state.
         The values obtained within the independent-electron approximation (to zeroth order in $1/Z$) 
         are given in terms of the function $A(\alpha Z)$ defined by Eq.~(\ref{eq:A_alphaZ}). 
         The interelectronic-interaction correction of first order in $1/Z$ is given in terms 
         of the function $B(\alpha Z)/Z$ defined by Eq.~(\ref{eq:B_alphaZ}). 
         The one-electron contribution is evaluated in the present work, 
         while the two-electron contribution is taken from Ref.~\cite{Malyshev:2019:012510}.
         }
         
\begin{tabular}{@{}
                l
                c@{\quad}
                S[table-format= 3.7]
                S[table-format=-2.8]
                S[table-format= 3.2]
                S[table-format= 2.8]
                S[table-format= 3.6]                
               }
               
\hline
\hline
   
   \multirow{2}{*}{$Z$}                               &
   \multirow{2}{*}{Approach}                          &
   \multicolumn{2}{c}{One-electron}                   &
   \multicolumn{2}{c}{Two-electron}                   &
   \multicolumn{1}{c}{\rule{0pt}{1.2em}\,\,\,Total}  \\
   \cmidrule(l{1.0em}r{1.1em}){3-4}
   \cmidrule(l{0.9em}r{1.2em}){5-6}
   \cmidrule(l{0.9em}r{0.1em}){7-7}    
   
                                                     &
                                                     &
   \multicolumn{1}{c}{$A$}                           &
   \multicolumn{1}{c}{$B/Z$}                         &
   \multicolumn{1}{c}{\,\,\,$A$}                     &
   \multicolumn{1}{c}{$B/Z$}                         &  
   \multicolumn{1}{c}{\,\,\,\,\,\,$A+B/Z$}     \\
        
\hline   
                       
  \multirow{2}{*}{  5}  &  QED \rule{0pt}{2.6ex}  &       1.000133   &       -0.125270   &             0.0   &        0.026735   &       0.901597   \\ 
                        &  $H_{\rm  MS}$  &       0.999999   &       -0.125244   &             0.0   &        0.026731   &       0.901486   \\[1.5mm] 

  \multirow{2}{*}{ 10}  &  QED            &       1.000937   &       -0.063080   &             0.0   &        0.013486   &       0.951343   \\ 
                        &  $H_{\rm  MS}$  &       0.999990   &       -0.062987   &             0.0   &        0.013473   &       0.950476   \\[1.5mm] 

  \multirow{2}{*}{ 20}  &  QED            &       1.006497   &       -0.032542   &             0.0   &        0.006988   &       0.980943   \\ 
                        &  $H_{\rm  MS}$  &       0.999907   &       -0.032220   &             0.0   &        0.006944   &       0.974631   \\[1.5mm] 

  \multirow{2}{*}{ 30}  &  QED            &       1.020211   &       -0.022964   &             0.0   &        0.004946   &       1.002193   \\ 
                        &  $H_{\rm  MS}$  &       0.999597   &       -0.022277   &             0.0   &        0.004844   &       0.982163   \\[1.5mm] 

  \multirow{2}{*}{ 40}  &  QED            &       1.045879   &       -0.018751   &             0.0   &        0.004031   &       1.031159   \\ 
                        &  $H_{\rm  MS}$  &       0.998745   &       -0.017532   &             0.0   &        0.003838   &       0.985051   \\[1.5mm] 

  \multirow{2}{*}{ 50}  &  QED            &       1.088554   &       -0.016842   &             0.0   &        0.003587   &       1.075299   \\ 
                        &  $H_{\rm  MS}$  &       0.996618   &       -0.014853   &             0.0   &        0.003256   &       0.985021   \\[1.5mm] 

  \multirow{2}{*}{ 60}  &  QED            &       1.155941   &       -0.016308   &             0.0   &        0.003403   &       1.143036   \\ 
                        &  $H_{\rm  MS}$  &       0.991803   &       -0.013186   &             0.0   &        0.002874   &       0.981491   \\[1.5mm] 

  \multirow{2}{*}{ 70}  &  QED            &       1.260160   &       -0.016889   &             0.0   &        0.003401   &       1.246672   \\ 
                        &  $H_{\rm  MS}$  &       0.980589   &       -0.012044   &             0.0   &        0.002590   &       0.971134   \\[1.5mm] 

  \multirow{2}{*}{ 80}  &  QED            &       1.423296   &       -0.018713   &             0.0   &        0.003559   &       1.408143   \\ 
                        &  $H_{\rm  MS}$  &       0.956496   &       -0.011133   &             0.0   &        0.002346   &       0.947709   \\[1.5mm] 

  \multirow{2}{*}{ 90}  &  QED            &       1.681978   &       -0.022263   &             0.0   &        0.003896   &       1.663612   \\ 
                        &  $H_{\rm  MS}$  &       0.900953   &       -0.010088   &             0.0   &        0.002093   &       0.892958   \\[1.5mm] 

  \multirow{2}{*}{ 92}  &  QED            &       1.750397   &       -0.023262   &             0.0   &        0.003990   &       1.731125   \\ 
                        &  $H_{\rm  MS}$  &       0.883145   &       -0.009816   &             0.0   &        0.002036   &       0.875365   \\[1.5mm] 

  \multirow{2}{*}{ 95}  &  QED            &       1.867993   &       -0.025016   &             0.0   &        0.004149   &       1.847127   \\ 
                        &  $H_{\rm  MS}$  &       0.850865   &       -0.009346   &             0.0   &        0.001943   &       0.843462   \\[1.5mm] 

  \multirow{2}{*}{100}  &  QED            &       2.114564   &       -0.028806   &             0.0   &        0.004476   &       2.090235   \\ 
                        &  $H_{\rm  MS}$  &       0.776885   &       -0.008318   &             0.0   &        0.001763   &       0.770329   \\[0.5mm]

\hline
\hline

\end{tabular}%

\end{table*}

\begin{table*}[t]
\centering

\renewcommand{\arraystretch}{1.1}

\caption{\label{tab:01:bind_1s1s2s} 
         The nuclear recoil contribution to the binding energy of the $1s^22s$ state.
         The values obtained within the independent-electron approximation (to zeroth order in $1/Z$) 
         are given in terms of the function $A(\alpha Z)$ defined by Eq.~(\ref{eq:A_alphaZ}). 
         The interelectronic-interaction correction of first order in $1/Z$ is given in terms 
         of the function $B(\alpha Z)/Z$ defined by Eq.~(\ref{eq:B_alphaZ}). 
         The one-electron contribution is evaluated in the present work, 
         while the two-electron contribution is taken from Ref.~\cite{Malyshev:2019:012510}.
         }
         
\begin{tabular}{@{}
                l
                c@{\quad}
                S[table-format= 3.7]
                S[table-format=-2.8]
                S[table-format= 3.2]
                S[table-format= 2.8]
                S[table-format= 3.6]                
               }
               
\hline
\hline
   
   \multirow{2}{*}{$Z$}                               &
   \multirow{2}{*}{Approach}                          &
   \multicolumn{2}{c}{One-electron}                   &
   \multicolumn{2}{c}{Two-electron}                   &
   \multicolumn{1}{c}{\rule{0pt}{1.2em}\,\,\,Total}  \\
   \cmidrule(l{1.0em}r{1.1em}){3-4}
   \cmidrule(l{0.9em}r{1.2em}){5-6}
   \cmidrule(l{0.9em}r{0.1em}){7-7}    
   
                                                     &
                                                     &
   \multicolumn{1}{c}{$A$}                           &
   \multicolumn{1}{c}{$B/Z$}                         &
   \multicolumn{1}{c}{\,\,\,$A$}                     &
   \multicolumn{1}{c}{$B/Z$}                         &  
   \multicolumn{1}{c}{\,\,\,\,\,\,$A+B/Z$}     \\
        
\hline   
                       
  \multirow{2}{*}{  5}  &  QED \rule{0pt}{2.6ex}  &       1.125184   &       -0.204919   &             0.0   &        0.031254   &       0.951519   \\ 
                        &  $H_{\rm  MS}$  &       1.125041   &       -0.204885   &             0.0   &        0.031250   &       0.951406   \\[1.5mm] 

  \multirow{2}{*}{ 10}  &  QED            &       1.126174   &       -0.103049   &             0.0   &        0.015782   &       1.038907   \\ 
                        &  $H_{\rm  MS}$  &       1.125157   &       -0.102927   &             0.0   &        0.015768   &       1.037998   \\[1.5mm] 

  \multirow{2}{*}{ 20}  &  QED            &       1.132673   &       -0.052858   &             0.0   &        0.008209   &       1.088024   \\ 
                        &  $H_{\rm  MS}$  &       1.125574   &       -0.052430   &             0.0   &        0.008162   &       1.081306   \\[1.5mm] 

  \multirow{2}{*}{ 30}  &  QED            &       1.148368   &       -0.036940   &             0.0   &        0.005841   &       1.117269   \\ 
                        &  $H_{\rm  MS}$  &       1.126105   &       -0.036022   &             0.0   &        0.005732   &       1.095815   \\[1.5mm] 

  \multirow{2}{*}{ 40}  &  QED            &       1.177480   &       -0.029770   &             0.0   &        0.004792   &       1.152501   \\ 
                        &  $H_{\rm  MS}$  &       1.126443   &       -0.028132   &             0.0   &        0.004581   &       1.102892   \\[1.5mm] 

  \multirow{2}{*}{ 50}  &  QED            &       1.225681   &       -0.026328   &             0.0   &        0.004295   &       1.203648   \\ 
                        &  $H_{\rm  MS}$  &       1.125841   &       -0.023641   &             0.0   &        0.003927   &       1.106127   \\[1.5mm] 

  \multirow{2}{*}{ 60}  &  QED            &       1.301675   &       -0.025069   &             0.0   &        0.004103   &       1.280709   \\ 
                        &  $H_{\rm  MS}$  &       1.122836   &       -0.020832   &             0.0   &        0.003505   &       1.105510   \\[1.5mm] 

  \multirow{2}{*}{ 70}  &  QED            &       1.419249   &       -0.025535   &             0.0   &        0.004129   &       1.397843   \\ 
                        &  $H_{\rm  MS}$  &       1.113475   &       -0.018924   &             0.0   &        0.003199   &       1.097750   \\[1.5mm] 

  \multirow{2}{*}{ 80}  &  QED            &       1.603600   &       -0.027874   &             0.0   &        0.004350   &       1.580076   \\ 
                        &  $H_{\rm  MS}$  &       1.090736   &       -0.017458   &             0.0   &        0.002940   &       1.076217   \\[1.5mm] 

  \multirow{2}{*}{ 90}  &  QED            &       1.897002   &       -0.032780   &             0.0   &        0.004793   &       1.869014   \\ 
                        &  $H_{\rm  MS}$  &       1.034184   &       -0.015901   &             0.0   &        0.002666   &       1.020949   \\[1.5mm] 

  \multirow{2}{*}{ 92}  &  QED            &       1.974823   &       -0.034191   &             0.0   &        0.004914   &       1.945547   \\ 
                        &  $H_{\rm  MS}$  &       1.015556   &       -0.015512   &             0.0   &        0.002603   &       1.002647   \\[1.5mm] 

  \multirow{2}{*}{ 95}  &  QED            &       2.108737   &       -0.036684   &             0.0   &        0.005120   &       2.077172   \\ 
                        &  $H_{\rm  MS}$  &       0.981468   &       -0.014846   &             0.0   &        0.002500   &       0.969122   \\[1.5mm] 

  \multirow{2}{*}{100}  &  QED            &       2.390112   &       -0.042124   &             0.0   &        0.005542   &       2.353531   \\ 
                        &  $H_{\rm  MS}$  &       0.902234   &       -0.013402   &             0.0   &        0.002294   &       0.891126   \\[0.5mm]

\hline
\hline

\end{tabular}%

\end{table*}

\begin{table*}[t]
\centering

\renewcommand{\arraystretch}{1.1}

\caption{\label{tab:01:bind_1s1s2p1} 
         The nuclear recoil contribution to the binding energy of the $1s^22p_{1/2}$ state.
         The values obtained within the independent-electron approximation (to zeroth order in $1/Z$) 
         are given in terms of the function $A(\alpha Z)$ defined by Eq.~(\ref{eq:A_alphaZ}). 
         The interelectronic-interaction correction of first order in $1/Z$ is given in terms 
         of the function $B(\alpha Z)/Z$ defined by Eq.~(\ref{eq:B_alphaZ}). 
         The one-electron contribution is evaluated in the present work, 
         while the two-electron contribution is taken from Ref.~\cite{Malyshev:2019:012510}.
         }
         
\begin{tabular}{@{}
                l
                c
                S[table-format= 3.7]
                S[table-format=-2.8]
                S[table-format=-3.7]
                S[table-format= 2.8]
                S[table-format= 3.6]                
               }
               
\hline
\hline
   
   \multirow{2}{*}{$Z$}                               &
   \multirow{2}{*}{Approach}                          &
   \multicolumn{2}{c}{One-electron}                   &
   \multicolumn{2}{c}{Two-electron}                   &
   \multicolumn{1}{c}{\rule{0pt}{1.2em}\,\,\,Total}  \\
   \cmidrule(l{1.0em}r{1.1em}){3-4}
   \cmidrule(l{1.2em}r{1.2em}){5-6}
   \cmidrule(l{0.9em}r{0.1em}){7-7} 
   
                                                     &
                                                     &
   \multicolumn{1}{c}{$A$}                           &
   \multicolumn{1}{c}{$B/Z$}                         &
   \multicolumn{1}{c}{\,\,\,$A$}                     &
   \multicolumn{1}{c}{$B/Z$\,\,\,}                   &   
   \multicolumn{1}{c}{\,\,\,\,\,\,$A+B/Z$}           \\
        
\hline   
                       
  \multirow{2}{*}{  5}  &  QED \rule{0pt}{2.6ex}  &       1.125174   &       -0.219111   &       -0.077986   &        0.088710   &       0.916786   \\ 
                        &  $H_{\rm  MS}$  &       1.125041   &       -0.219084   &       -0.077986   &        0.088706   &       0.916676   \\[1.5mm] 

  \multirow{2}{*}{ 10}  &  QED            &       1.126100   &       -0.110207   &       -0.077835   &        0.044519   &       0.982578   \\ 
                        &  $H_{\rm  MS}$  &       1.125157   &       -0.110109   &       -0.077833   &        0.044505   &       0.981720   \\[1.5mm] 

  \multirow{2}{*}{ 20}  &  QED            &       1.132153   &       -0.056533   &       -0.077225   &        0.022595   &       1.020990   \\ 
                        &  $H_{\rm  MS}$  &       1.125580   &       -0.056188   &       -0.077196   &        0.022542   &       1.014739   \\[1.5mm] 

  \multirow{2}{*}{ 30}  &  QED            &       1.146728   &       -0.039464   &       -0.076199   &        0.015449   &       1.046514   \\ 
                        &  $H_{\rm  MS}$  &       1.126131   &       -0.038717   &       -0.076046   &        0.015315   &       1.026684   \\[1.5mm] 

  \multirow{2}{*}{ 40}  &  QED            &       1.173729   &       -0.031708   &       -0.074741   &        0.012011   &       1.079291   \\ 
                        &  $H_{\rm  MS}$  &       1.126526   &       -0.030361   &       -0.074234   &        0.011734   &       1.033666   \\[1.5mm] 

  \multirow{2}{*}{ 50}  &  QED            &       1.218400   &       -0.027889   &       -0.072820   &        0.010076   &       1.127767   \\ 
                        &  $H_{\rm  MS}$  &       1.126073   &       -0.025651   &       -0.071506   &        0.009563   &       1.038480   \\[1.5mm] 

  \multirow{2}{*}{ 60}  &  QED            &       1.288793   &       -0.026341   &       -0.070388   &        0.008916   &       1.200980   \\ 
                        &  $H_{\rm  MS}$  &       1.123422   &       -0.022760   &       -0.067442   &        0.008034   &       1.041254   \\[1.5mm] 

  \multirow{2}{*}{ 70}  &  QED            &       1.397653   &       -0.026548   &       -0.067367   &        0.008234   &       1.311973   \\ 
                        &  $H_{\rm  MS}$  &       1.114927   &       -0.020872   &       -0.061327   &        0.006777   &       1.039506   \\[1.5mm] 

  \multirow{2}{*}{ 80}  &  QED            &       1.568287   &       -0.028627   &       -0.063632   &        0.007903   &       1.483931   \\ 
                        &  $H_{\rm  MS}$  &       1.094163   &       -0.019528   &       -0.051886   &        0.005537   &       1.028286   \\[1.5mm] 

  \multirow{2}{*}{ 90}  &  QED            &       1.839863   &       -0.033286   &       -0.058988   &        0.007886   &       1.755474   \\ 
                        &  $H_{\rm  MS}$  &       1.042421   &       -0.018239   &       -0.036694   &        0.004028   &       0.991517   \\[1.5mm] 

  \multirow{2}{*}{ 92}  &  QED            &       1.911923   &       -0.034657   &       -0.057926   &        0.007923   &       1.827262   \\ 
                        &  $H_{\rm  MS}$  &       1.025379   &       -0.017927   &       -0.032597   &        0.003658   &       0.978513   \\[1.5mm] 

  \multirow{2}{*}{ 95}  &  QED            &       2.035959   &       -0.037104   &       -0.056235   &        0.008007   &       1.950628   \\ 
                        &  $H_{\rm  MS}$  &       0.994205   &       -0.017395   &       -0.025536   &        0.003036   &       0.954310   \\[1.5mm] 

  \multirow{2}{*}{100}  &  QED            &       2.296765   &       -0.042522   &       -0.053123   &        0.008235   &       2.209355   \\ 
                        &  $H_{\rm  MS}$  &       0.921775   &       -0.016226   &       -0.010645   &        0.001757   &       0.896661   \\[0.5mm]

\hline
\hline

\end{tabular}%

\end{table*}


\begin{figure}
\begin{center}
\includegraphics[width=\columnwidth]{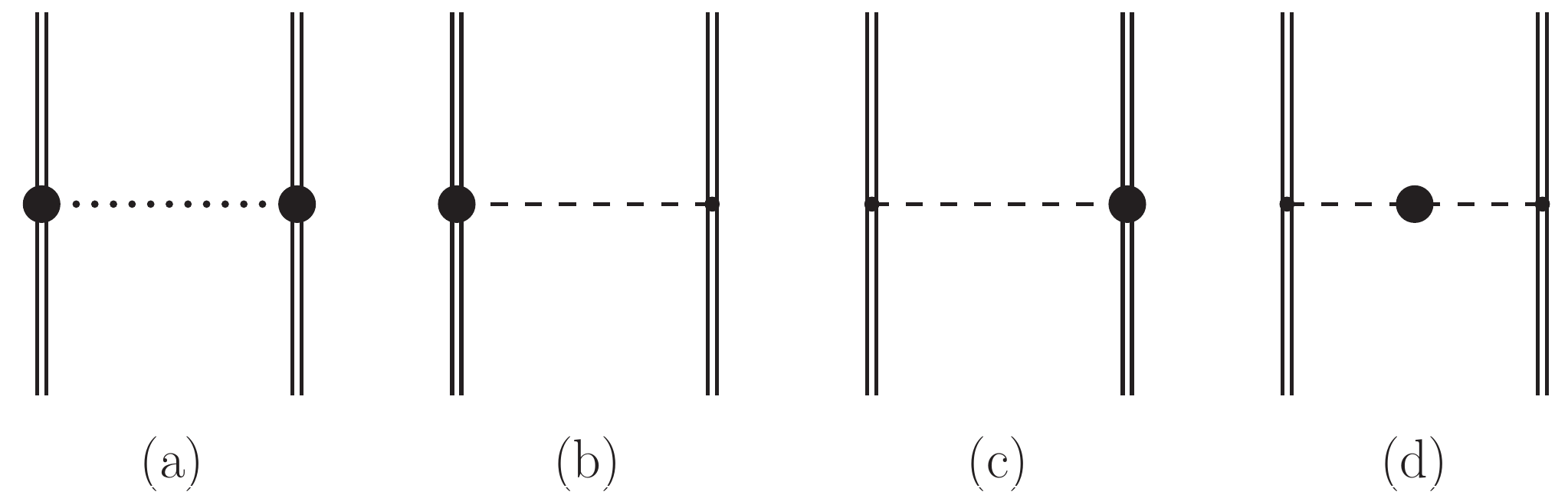}
\caption{\label{fig:recoil_2el}
Two-electron nuclear recoil diagrams to zeroth order in $1/Z$: the Coulomb~(a), one-transverse~(b) and (c), and two-transverse~(d) contributions.}
\end{center}
\end{figure}

\begin{figure}
\begin{center}
\includegraphics[width=\columnwidth]{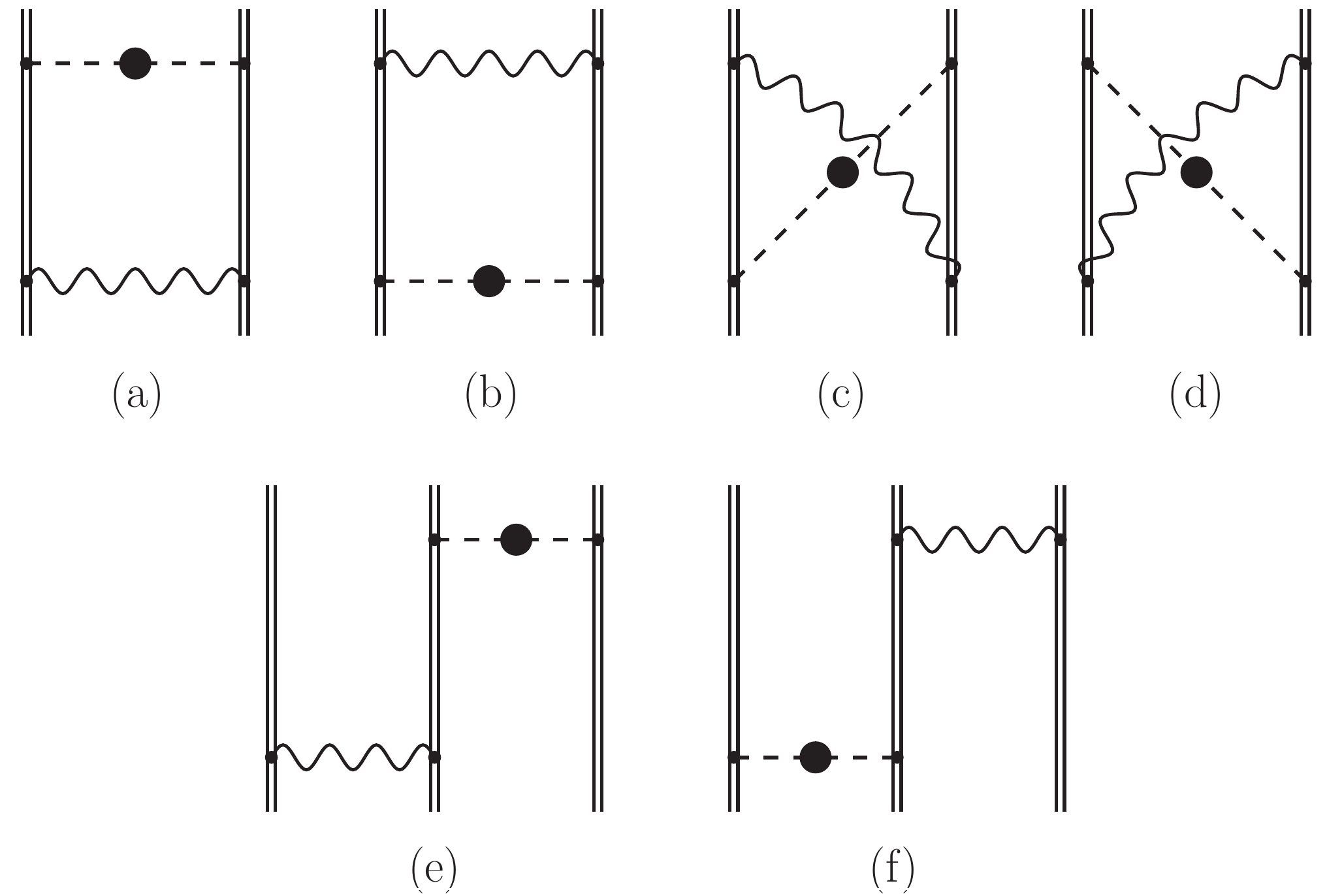}
\caption{\label{fig:recoil_2el_IntEl}
The interelectronic-interaction correction to the two-electron two-transverse-photon contribution to the nuclear recoil effect. The analogous diagrams with the Coulomb and one-transverse-photon recoil interactions have to be taken into account as well.
}
\end{center}
\end{figure}

As noted in Sec.~\ref{sec:0}, the one-electron part of the nuclear recoil effect, which is the major focus of the present work, has to be combined with the two-electron part in order to complete the rigorous consideration to first order in $1/Z$. The two-electron nuclear recoil contribution was studied nonperturbatively (in $\alpha Z$) in our recent paper~\cite{Malyshev:2019:012510}. Within the independent-electron approximation, the two-electron contribution is described by the diagrams depicted in Fig.~\ref{fig:recoil_2el}. The interelectronic-interaction correction to the two-electron part of the nuclear recoil effect is given by the diagrams displayed in Fig.~\ref{fig:recoil_2el_IntEl}. The notations for the diagram technique are the same as in Figs.~\ref{fig:recoil_1el} and \ref{fig:recoil_1el_IntEl}. We note that the diagrams analogous to those in Fig.~\ref{fig:recoil_2el_IntEl} with the two-transverse-photon interaction replaced with the Coulomb and one-transverse-photon interactions have to be taken into account as well. 

\begin{figure}
\begin{center}
\includegraphics[width=\columnwidth]{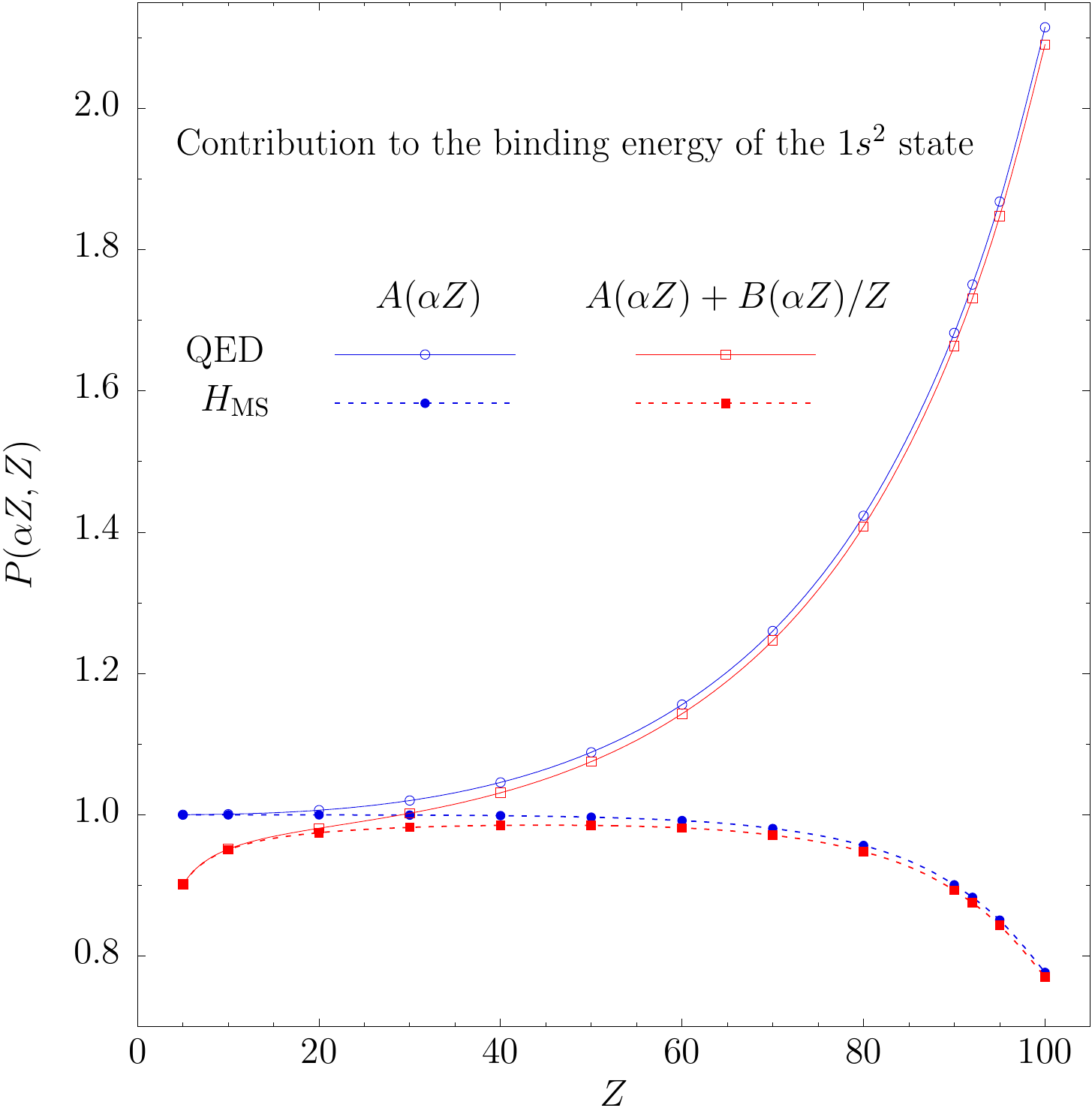}
\caption{\label{fig:01:bind_1s1s}
The nuclear recoil effect on the binding energy of the $1s^2$ state to first order in $1/Z$. The solid lines stand for the results of the QED calculations to all orders in $\alpha Z$ while the dashed lines correspond to the calculations based on the mass shift (MS) operator given by Eq.~(\ref{eq:NMS+SMS}). The contributions of zeroth order in $1/Z$, $P_{[0]}(\alpha Z)=A(\alpha Z)$, and the sums of zeroth and first orders in $1/Z$, $P_{[0,1]}(\alpha Z, Z)=A(\alpha Z) + B(\alpha Z)/Z$, are shown with blue (circles) and red (squares) lines, respectively.}
\end{center}
\end{figure}

\begin{figure}
\begin{center}
\includegraphics[width=\columnwidth]{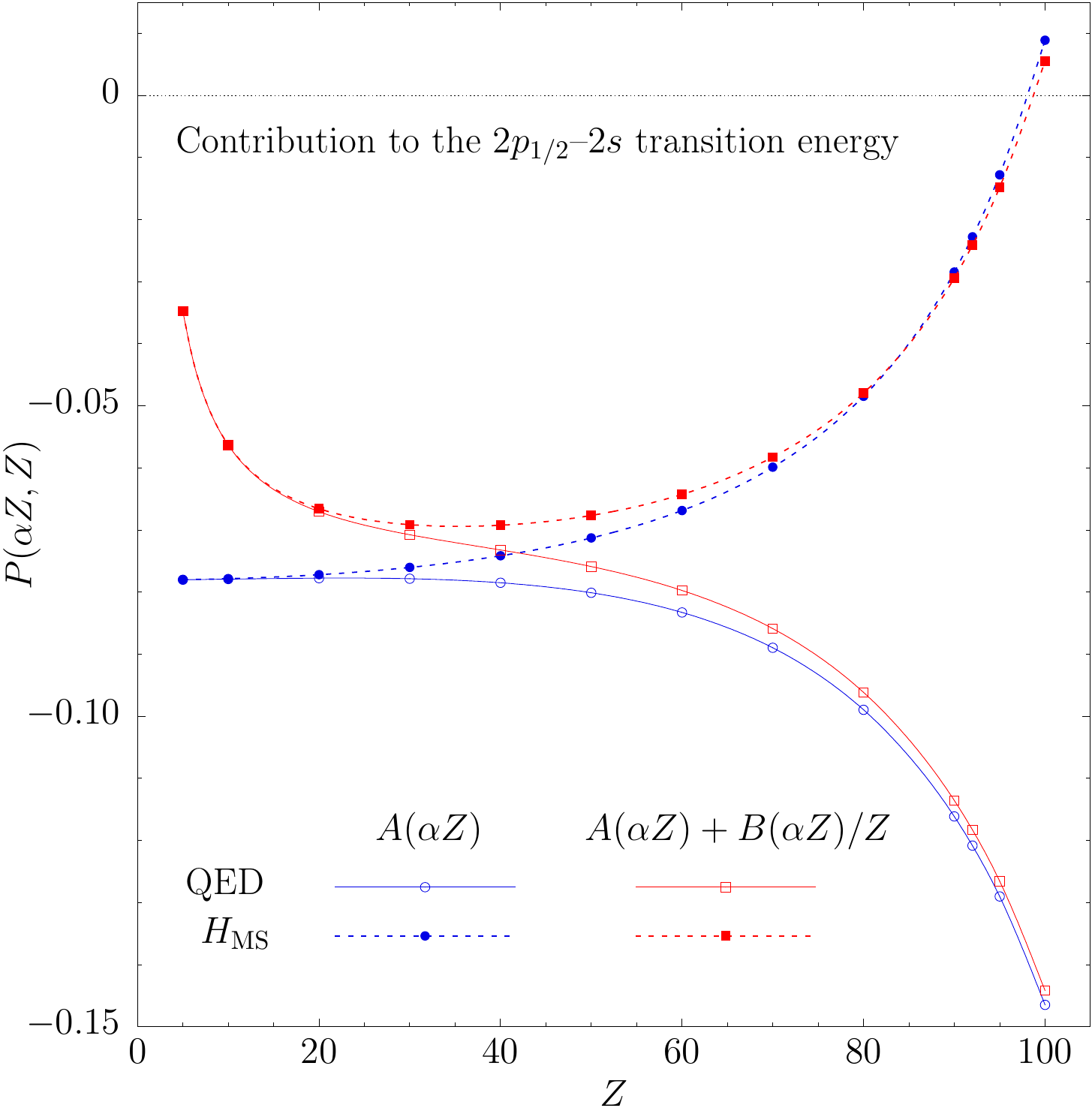}
\caption{\label{fig:01:tr_2p1_2s}
The nuclear recoil effect on the $2p_{1/2}$--$2s$ transition energy in Li-like ions to first order in $1/Z$. Notations are the same as in Fig.~\ref{fig:01:bind_1s1s}.}
\end{center}
\end{figure}

In Tables~\ref{tab:01:bind_1s1s}--\ref{tab:01:bind_1s1s2p1}, we summarize the data obtained to first order in $1/Z$ for the nuclear recoil effect on the binding energies of the $1s^2$, $1s^2 2s$, and $1s^2 2p_{1/2}$ states. The results for the zeroth- and first-order (in $1/Z$) contributions are presented in terms of the functions $A(\alpha Z)$ and $B(\alpha Z)/Z$, respectively. The one-electron part of the nuclear recoil effect is calculated in the present work. To zeroth order in $1/Z$, the one-electron contribution to binding energy is obtained by summing the values from Table~\ref{tab:0:all_1el_contrib1s_2s_2p1} for all the electrons involved. The interelectronic-interaction correction to the one-electron part is taken from Tables~\ref{tab:1:bind_1s1s}--\ref{tab:1:bind_1s1s2p1}. For the two-electron contribution the data from Ref.~\cite{Malyshev:2019:012510} are used. We note that for the $1s^2$ and $1s^2 2s$ states the two-electron part vanishes identically within the independent-electron approximation. For each state, the sum of the zeroth- and first-order contributions, $A(\alpha Z)+B(\alpha Z)/Z$, is presented in the last column. As above, in Tables~\ref{tab:01:bind_1s1s}--\ref{tab:01:bind_1s1s2p1} we compare the results obtained by means of the rigorous QED approach and within the Breit approximation via the MS operator (\ref{eq:NMS+SMS}). For illustrative purposes, the data for the binding energy of the $1s^2$ state from Table~\ref{tab:01:bind_1s1s} and the corresponding data for the $2p_{1/2}$--$2s$ transition energy in Li-like ions are plotted in Figs.~\ref{fig:01:bind_1s1s} and \ref{fig:01:tr_2p1_2s}, respectively. The data for the transition energy are obtained as the difference of the values presented in Tables~\ref{tab:01:bind_1s1s2p1} and \ref{tab:01:bind_1s1s2s}. In Figs.~\ref{fig:01:bind_1s1s} and \ref{fig:01:tr_2p1_2s}, the dashed lines correspond to the calculations within the Breit approximation using the MS operator (\ref{eq:NMS+SMS}), while the solid lines represent the QED results valid to all orders in $\alpha Z$. The contributions within the independent-electron approximation, $A(\alpha Z)$, are shown with the blue lines with circles. The sum of the zeroth and first orders in $1/Z$, $A(\alpha Z)+B(\alpha Z)/Z$, are given with the red lines with squares. There is no doubt, that the convergence of the $1/Z$-perturbation theory may be slow for low-$Z$ ions. For this reason, the results presented in Tables~\ref{tab:01:bind_1s1s}--\ref{tab:01:bind_1s1s2p1} and Figs.~\ref{fig:01:bind_1s1s} and \ref{fig:01:tr_2p1_2s} should not be considered as the final ones for low- and middle-$Z$ systems; the contribution of the higher orders in $1/Z$ can be significant (see the discussion below). Nevertheless, these data yield insights into the state-of-the-art QED calculations of the nuclear recoil effect to all orders in $\alpha Z$ and give the indication of how different terms relate to each other. 

We stress that the interelectronic-interaction correction under consideration becomes particularly important when a cancellation of the zeroth-order contributions occurs. For instance, the one-electron contribution for the $1s^2$ core cancels in the $2p_{1/2}$--$2s$ transition in Li-like ions within the independent-electron approximation. As a result, the nontrivial QED contributions of zeroth and first orders in $1/Z$ are of comparable magnitude for low- and middle-$Z$ ions for this transition. In this regard, one can expect even stronger cancellation of the leading-order contributions in the case of the $2p_{3/2}$--$2p_{1/2}$ transition in B-like ions; see the related discussion for the QED contribution to the field shift in Ref.~\cite{Zubova:2016:052502}. In addition, the \textit{ab initio} treatment of the electron-electron interaction correction to all orders in $\alpha Z$ may even change the sign of the correction. Indeed, one can see that the solid lines in Fig.~\ref{fig:01:tr_2p1_2s} do not cross each other in contrast to the dashed ones. All this leads to the conclusion that the high-precision calculations of the nuclear recoil effect need to take into account the QED contribution beyond the independent-electron approximation.

\subsection{Mass shift of binding and transition energies}

As noted above, in order to obtain accurate theoretical predictions for the mass shift of binding and transition energies one has to account for the second- and higher-order electron-electron interaction corrections to the nuclear recoil effect as well. In the present work, we evaluate these contributions within the lowest-order relativistic approximation by employing the MS operator (\ref{eq:NMS+SMS}) and  the Dirac-Coulomb-Breit Hamiltonian. The calculations are performed by means of two independent methods. First, we have calculated the expectation value of the MS operator with the many-electron wave function obtained by the configuration-interaction method in the basis of the Dirac-Sturm orbitals~\cite{Bratzev:1977:173, Tupitsyn:2003:022511}; see also Ref.~\cite{Kaygorodov:2019:032505}. The desired higher-order correction has been extracted by subtracting the zeroth- and first-order contributions evaluated with the same basis set. Second, we have employed the recursive formulation of the perturbation theory~\cite{Glazov:2017:46} in order to directly access the required higher-order correction. This method has been applied already for evaluation of the higher-order nuclear recoil contributions to the ionization energies in boronlike ions \cite{Malyshev:2017:022512} and to the bound-electron $g$ factor in lithiumlike~\cite{Shabaev:2017:263001,Shabaev:2018:032512} and boronlike~\cite{Glazov:2020:012515} ions. The results of both independent approaches are found in good agreement with each other. 

\begin{table*}[t]
\centering

\renewcommand{\arraystretch}{1.0}

\caption{\label{tab:2plus} 
         The interelectronic-interaction correction of second and higher orders in $1/Z$
         to binding energies of the $1s^2$, $1s^2 2s$, and $1s^2 2p_{1/2}$ states within the Breit approximation
         expressed in terms of the dimensionless function $C(\alpha Z,Z)$ defined by Eq.~(\ref{eq:C_alphaZ_Z}). 
         }
         
\resizebox{\textwidth}{!}{%
\begin{tabular}{@{}
                l
                S[table-format= 3.4(2)]
                S[table-format=-2.4(2)]
                S[table-format= 2.5(2)]
                S[table-format= 3.4(2)]
                S[table-format=-2.4(2)]  
                S[table-format= 2.5(2)]
                S[table-format= 3.4(2)]
                S[table-format=-2.4(2)]
                S[table-format= 2.4(2)]                
               }
               
\hline
\hline
   
   \multirow{2}{*}{$Z$}                                    &
   \multicolumn{3}{c}{$1s^2$}                              &
   \multicolumn{3}{c}{$1s^2 2s$}                           &
   \multicolumn{3}{c}{\rule{0pt}{1.2em}$1s^2 2p_{1/2}$}  \\
   \cmidrule(l{1.0em}r{1.1em}){2-4}
   \cmidrule(l{0.9em}r{1.2em}){5-7}
   \cmidrule(l{0.9em}r{0.1em}){8-10}    
   
                                                    &
   \multicolumn{1}{c}{$C_{\rm 1el}$}               &
   \multicolumn{1}{c}{$C_{\rm 2el}$}               &
   \multicolumn{1}{c}{$C_{\rm tot}$}               &
   \multicolumn{1}{c}{$C_{\rm 1el}$}               &
   \multicolumn{1}{c}{$C_{\rm 2el}$}               &
   \multicolumn{1}{c}{$C_{\rm tot}$}               &
   \multicolumn{1}{c}{$C_{\rm 1el}$}               &
   \multicolumn{1}{c}{$C_{\rm 2el}$}               &
   \multicolumn{1}{c}{$C_{\rm tot}$}               \\   
        
\hline   
                       
    5 \rule{0pt}{2.6ex} &      0.1580(2)  &     -0.1145(3)  &      0.0434(2)  &      0.4167(2)  &     -0.1741(3)  &      0.2426(2)  &      0.5513(2)  &     -0.3497(6)  &      0.2015(4) \\[1mm] 

   10                      &      0.1656(2)  &     -0.1196(5)  &      0.0460(3)  &      0.4233(2)  &     -0.1834(5)  &      0.2399(3)  &      0.5546(2)  &     -0.3445(8)  &      0.2101(5) \\[1mm] 

   20                      &      0.1930(5)  &    -0.1328(10)  &      0.0601(5)  &      0.4624(5)  &    -0.2039(11)  &      0.2584(6)  &      0.6070(6)  &    -0.3688(15)  &      0.2382(9) \\[1mm] 

   30                      &      0.2396(7)  &    -0.1534(13)  &      0.0862(6)  &      0.5322(8)  &    -0.2351(14)  &      0.2971(6)  &      0.7050(9)  &    -0.4144(18)  &      0.2906(9) \\[1mm] 

   40                      &     0.3045(11)  &    -0.1813(17)  &      0.1232(7)  &     0.6315(13)  &    -0.2772(18)  &      0.3543(7)  &     0.8482(14)  &    -0.4768(23)  &     0.3713(10) \\[1mm] 

   50                      &     0.3900(17)  &    -0.2163(21)  &     0.1737(10)  &     0.7650(19)  &    -0.3306(23)  &     0.4344(11)  &     1.0439(20)  &    -0.5532(29)  &     0.4906(13) \\[1mm] 

   60                      &     0.4985(27)  &    -0.2582(27)  &     0.2403(17)  &     0.9376(31)  &    -0.3945(30)  &     0.5430(20)  &     1.3022(31)  &    -0.6372(36)  &     0.6650(20) \\[1mm] 

   70                      &     0.6322(42)  &    -0.3055(35)  &     0.3267(30)  &     1.1546(49)  &    -0.4674(38)  &     0.6872(35)  &     1.6362(47)  &    -0.7142(45)  &     0.9221(32) \\[1mm] 

   80                      &     0.7902(66)  &    -0.3549(44)  &     0.4353(50)  &     1.4169(77)  &    -0.5444(49)  &     0.8725(60)  &     2.0582(72)  &    -0.7509(56)  &     1.3073(53) \\[1mm] 

   90                      &      0.954(10)  &    -0.3966(57)  &     0.5569(80)  &      1.696(12)  &    -0.6116(64)  &     1.0840(95)  &      2.552(11)  &    -0.6655(71)  &     1.8861(83) \\[1mm] 

   92                      &      0.982(11)  &    -0.4025(60)  &     0.5793(87)  &      1.745(13)  &    -0.6214(67)  &      1.124(10)  &      2.651(12)  &    -0.6194(75)  &     2.0314(90) \\[1mm] 

   95                      &      1.018(12)  &    -0.4080(64)  &      0.610(10)  &      1.809(14)  &    -0.6316(73)  &      1.177(12)  &      2.793(13)  &    -0.5202(80)  &      2.272(10) \\[1mm] 

  100                      &      1.046(15)  &    -0.4046(74)  &      0.641(12)  &      1.864(18)  &    -0.6301(84)  &      1.234(14)  &      2.979(16)  &    -0.2411(90)  &      2.738(13) \\[1mm]

\hline
\hline

\end{tabular}%
}

\end{table*}

\begin{table*}[t]
\centering

\renewcommand{\arraystretch}{1.0}

\caption{\label{tab:total} 
         The mass shifts of the binding energies of the $1s^2$, $1s^2 2s$, and $1s^2 2p_{1/2}$ states
         and the mass shift of the $2p_{1/2}$--$2s$ transition energy in Li-like ions in terms of the dimensionless function
         $P(\alpha Z,Z)$ defined by Eq.~(\ref{eq:P_alphaZ_Z}) and the $K$ factor (in~$\rm eV\!\cdot\! amu$) 
         defined by Eq.~(\ref{eq:Kfactor}). 
         }
         
\resizebox{\textwidth}{!}{%
\begin{tabular}{@{}
                l
                S[table-format= 3.5(2),group-separator=]
                S[table-format= 3.6(2),group-separator=]
                S[table-format= 3.5(2),group-separator=]
                S[table-format= 3.6(2),group-separator=]
                S[table-format= 3.5(2),group-separator=]
                S[table-format= 3.6(2),group-separator=]
                S[table-format=-3.6(2),group-separator=]
                S[table-format=-2.6(1),group-separator=]
                @{}
               }
               
\hline
\hline
   
   \multirow{2}{*}{$Z$}                                    &
   \multicolumn{2}{c}{\rule{0pt}{1.2em}$1s^2$}             &
   \multicolumn{2}{c}{$1s^22s$}                            &
   \multicolumn{2}{c}{$1s^22p_{1/2}$}                      &
   \multicolumn{2}{c}{$2p_{1/2}$--$2s$}                  \\
   \cmidrule(l{1.0em}r{1.0em}){2-3}
   \cmidrule(l{1.0em}r{1.0em}){4-5}
   \cmidrule(l{1.0em}r{1.0em}){6-7} 
   \cmidrule(l{1.0em}r{0.8em}){8-9}
   
                                                &
   \multicolumn{1}{c}{$P$}                     &
   \multicolumn{1}{c}{$K$}                     &
   \multicolumn{1}{c}{$P$}                     &
   \multicolumn{1}{c}{$K$}                     &
   \multicolumn{1}{c}{$P$}                     &
   \multicolumn{1}{c}{$K$}                     &
   \multicolumn{1}{c}{$P$}                     &
   \multicolumn{1}{c}{$K$}                     \\

\hline   
                       
    5 \rule{0pt}{2.6ex} &     0.90334(1)  &    0.337116(3)  &     0.96122(1)  &    0.358719(3)  &     0.92485(2)  &     0.34514(1)  &  -0.036376(19)  &   -0.013575(7) \\[1.5mm] 

   10                      &     0.95180(1)  &     1.42082(2)  &     1.04131(2)  &     1.55442(2)  &     0.98468(1)  &     1.46989(2)  &   -0.056628(6)  &    -0.08453(1)   \\ 
                           &                 &                 &                 &                 &                 &                 &                 &    -0.08456(2) \\[1.5mm] 

   20                      &    0.98110(10)  &    5.85817(57)  &    1.08867(10)  &    6.50051(62)  &    1.02159(10)  &    6.09994(58)  &   -0.067085(8)  &    -0.40057(5) \\[1.5mm] 

   30                      &    1.00230(30)  &    13.4658(40)  &    1.11761(32)  &    15.0150(44)  &    1.04685(30)  &    14.0643(40)  &  -0.070762(24)  &   -0.95068(32)   \\ 
                           &                 &                 &                 &                 &                 &                 &                 &    -0.9533(16) \\[1.5mm] 

   40                      &    1.03130(69)  &     24.632(16)  &    1.15279(74)  &     27.533(18)  &    1.07958(69)  &     25.785(16)  &  -0.073203(57)  &    -1.7484(14) \\[1.5mm] 

   50                      &     1.0756(13)  &     40.140(50)  &     1.2041(15)  &     44.935(54)  &     1.1282(13)  &     42.103(50)  &   -0.07587(12)  &    -2.8316(43) \\[1.5mm] 

   60                      &     1.1438(24)  &      61.47(13)  &     1.2817(26)  &      68.88(14)  &     1.2019(24)  &      64.59(13)  &   -0.07975(22)  &     -4.286(12)   \\ 
                           &                 &                 &                 &                 &                 &                 &                 &     -4.334(29) \\[1.5mm] 

   70                      &     1.2490(41)  &      91.36(30)  &     1.4004(45)  &     102.43(33)  &     1.3144(42)  &      96.14(31)  &   -0.08599(39)  &     -6.290(28)   \\ 
                           &                 &                 &                 &                 &                 &                 &                 &       -6.39(6) \\[1.5mm] 

   80                      &     1.4142(74)  &     135.11(71)  &     1.5868(81)  &     151.60(77)  &     1.4902(75)  &     142.37(72)  &   -0.09656(70)  &     -9.225(67)   \\ 
                           &                 &                 &                 &                 &                 &                 &                 &      -9.40(12) \\[1.5mm] 

   90                      &      1.680(18)  &      203.1(22)  &      1.887(20)  &      228.2(24)  &      1.772(18)  &      214.3(22)  &    -0.1148(15)  &     -13.88(19)   \\ 
                           &                 &                 &                 &                 &                 &                 &                 &     -14.23(24) \\[1.5mm] 

   92                      &      1.751(23)  &      221.2(29)  &      1.967(25)  &      248.6(32)  &      1.847(23)  &      233.4(29)  &    -0.1198(19)  &     -15.14(24)   \\ 
                           &                 &                 &                 &                 &                 &                 &                 &     -15.44(27) \\[1.5mm] 

   95                      &      1.873(35)  &      252.4(47)  &      2.106(37)  &      283.8(50)  &      1.978(35)  &      266.4(47)  &    -0.1287(28)  &     -17.34(37) \\[1.5mm] 

  100                      &      2.132(75)  &        318(11)  &      2.400(81)  &        358(12)  &      2.252(75)  &        336(11)  &    -0.1477(58)  &     -22.05(86) \\[1.5mm]

\hline
\hline

\end{tabular}%
}

\end{table*}

The second- and higher-order (in $1/Z$) interelectronic-interaction corrections to the nuclear recoil effect on binding energies of the $1s^2$, $1s^2 2s$, and $1s^2 2p_{1/2}$ states are presented in Table~\ref{tab:2plus} in terms of the dimensionless function $C(\alpha Z,Z)$ defined according to
\begin{equation}
\label{eq:C_alphaZ_Z}
\Delta E^{(3+)} = \frac{m}{M} \frac{(\alpha Z)^2}{Z^2} C(\alpha Z, Z) \, mc^2 \, .
\end{equation} 
The one- and two-electron parts of the corresponding contribution are evaluated with the use of the NMS (\ref{eq:NMS}) and SMS (\ref{eq:SMS}) operators, respectively, and given explicitly. The uncertainties specified in Table~\ref{tab:2plus} correspond to the numerical errors only. They are obtained by analyzing the convergence of the results with respect to the number of the radial and angular basis-set functions. We note that the two-electron part is more sensitive to the correlation effects than the one-electron part. As a result, the corresponding uncertainty is generally bigger for low- and middle-$Z$ ions. On the other hand, there is a cancellation between the one- and two-electron contributions which allows us to obtain more accurate data for the total values.

Finally, in Table~\ref{tab:total} we compile the total theoretical predictions for the mass shifts of the following quantities: (i) the ground-state binding energy of He-like ions; (ii) the binding energy of the $1s^2 2s$ state; (iii) the binding energy of the $1s^2 2p_{1/2}$ state; (iv) the $2p_{1/2}$--$2s$ transition energy in Li-like ions. The results are expressed in terms of the dimensionless function $P(\alpha Z, Z)$ defined according to
\begin{equation}
\label{eq:P_alphaZ_Z}
\Delta E_{\rm rec} = \frac{m}{M} (\alpha Z)^2 P(\alpha Z, Z) \, mc^2 \, 
\end{equation} 
and the $K$ factor (in units of $\rm eV\!\cdot\! amu$) defined by
\begin{equation}
\label{eq:Kfactor}
\Delta E_{\rm rec} = \frac{K}{M} \, . 
\end{equation} 
The total theoretical predictions comprise the QED results for the zeroth-order, $A(\alpha Z)$, and first-order, $B(\alpha Z)/Z$, contributions from Tables~\ref{tab:0:all_1el_contrib1s_2s_2p1} and \ref{tab:1:bind_1s1s}--\ref{tab:1:bind_1s1s2p1}, respectively, as well as the higher-order correlation correction within the Breit approximation, $C(\alpha Z, Z)/Z^2$, from Table~\ref{tab:2plus}. In addition, within the independent-electron approximation we account for the correction $\delta A_{\rm Breit}^{\rm fns,1el}(\alpha Z)$, which determines the difference between the exact treatment of the nuclear size correction to the low-order one-electron nuclear recoil effect and its evaluation by the formula~(\ref{eq:dE_rec_L}) with the wave functions for the extended nucleus; see Ref.~\cite{Aleksandrov:2015:144004} and the discussion above. Therefore, the function $P(\alpha Z, Z)$ in Eq.~(\ref{eq:P_alphaZ_Z}) can be represented as follows
\begin{equation}
\label{eq:P_alphaZ_Z_sum}
P(\alpha Z, Z) = A(\alpha Z)  + \delta A_{\rm Breit}^{\rm fns,1el}(\alpha Z) 
               + \frac{B(\alpha Z)}{Z} + \frac{C(\alpha Z, Z)}{Z^2}  \, .
\end{equation} 
We note that the reduced-mass dependence in the Lamb shift also contributes to the nuclear recoil effect; see, e.g., the discussion in Ref.~\cite{Yerokhin:2015:033103} and references therein. This contribution and the uncertainty related with it are out of the scope of the present work. They have to be taken into account separately.

\begin{figure*}
\subfloat[]{%
\includegraphics[height=0.415\textheight]{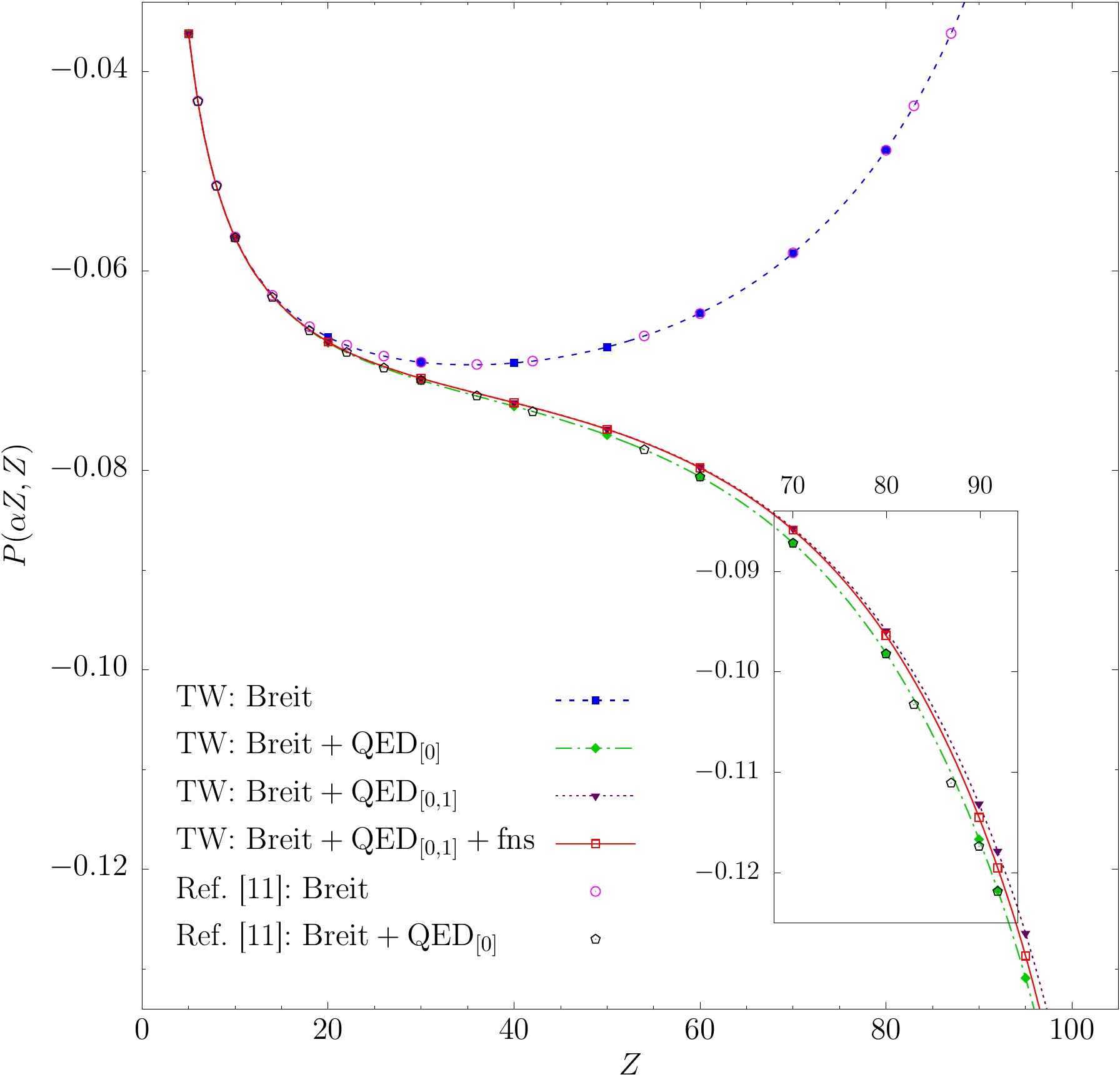}}
\hspace*{\fill}
\subfloat[]{%
\includegraphics[height=0.415\textheight]{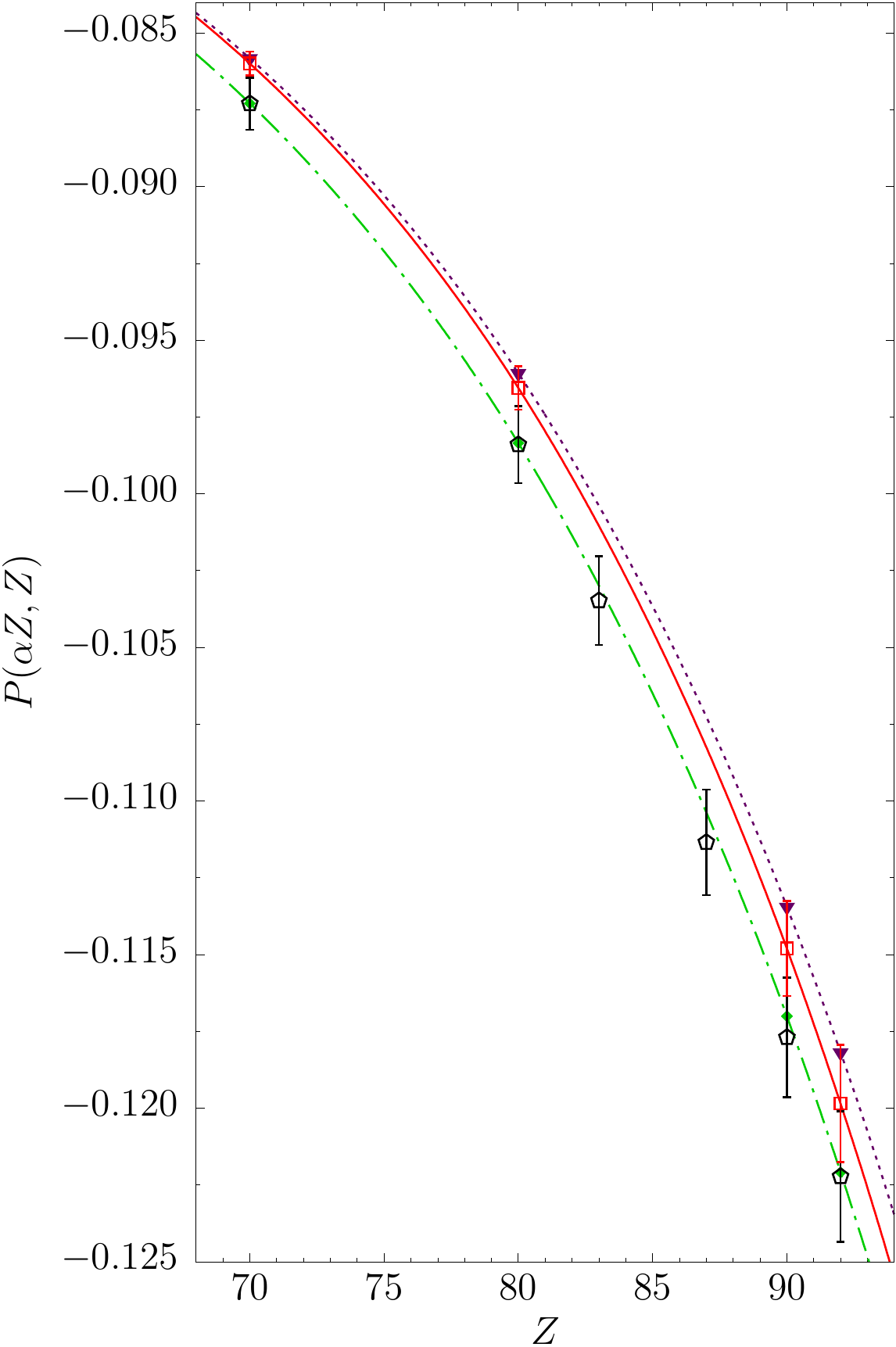}}
\caption{\label{fig:step_by_step:tr_2p1_2s}
(a)~The nuclear recoil effect on the $2p_{1/2}$--$2s$ transition energy in Li-like ions in terms of the function $P(\alpha Z, Z)$ defined by Eq.~(\ref{eq:P_alphaZ_Z}). The blue dashed line corresponds to the calculation performed by means of the mass shift operator (\ref{eq:NMS+SMS}) to all order in $1/Z$. The green dashed-dotted line includes the nontrivial QED contribution within the independent-electron approximation  ($\rm QED_{[0]}$). The violet dotted line accounts for the QED correction to first-order in $1/Z$ ($\rm QED_{[0,1]}$). The red solid line takes into account additionally the finite nuclear size (fns) correction~$\delta A_{\rm Breit}^{\rm fns,1el}(\alpha Z)$. The corresponding data from Ref.~\cite{Zubova:2014:062512} are shown with magenta circles and black diamonds. The error bars are not indicated. (b) The zoomed region for $Z=68$--$94$. The uncertainties of the present calculation and Ref.~\cite{Zubova:2014:062512} are shown.}
\end{figure*}

Besides the numerical uncertainties discussed above, there are several sources for the theoretical uncertainties shown in parentheses in Table~\ref{tab:total}. First of all, we take into account the uncertainty due to uncalculated radiative nuclear recoil correction. To this end, we multiply the nontrivial one-electron QED contribution (\ref{eq:dE_rec_H}) obtained within the independent-electron approximation by the factor of $2\alpha$. Second, we estimate the uncertainty due to the approximate treatment of the nuclear size correction to the nuclear recoil effect by using the prescription given in Refs.~\cite{Yerokhin:2015:033103, Malyshev:2018:085001}. Finally, all the uncertainties are combined by calculating their root sum square. 

In Table~\ref{tab:total}, we compare our total values for the mass shift of the $2p_{1/2}$--$2s$ transition energy in Li-like ions with the theoretical predictions from Ref.~\cite{Zubova:2014:062512}. One can see that the data from Ref.~\cite{Zubova:2014:062512} lie systematically lower. The more detailed comparison is performed in Figs.~\ref{fig:step_by_step:tr_2p1_2s}(a) and \ref{fig:step_by_step:tr_2p1_2s}(b) [Fig.~\ref{fig:step_by_step:tr_2p1_2s}(b) provides the zoomed version of Fig.~\ref{fig:step_by_step:tr_2p1_2s}(a) which corresponds to the high-$Z$ region ($Z=68$--$94$)]. The four lines labeled with TW in Fig.~\ref{fig:step_by_step:tr_2p1_2s} represent our data obtained by successive accounting for the different contributions. The blue dashed line displays the results calculated by employing the MS operator (\ref{eq:NMS+SMS}) and treating the correlation effects to all orders in $1/Z$ within the Breit approximation. The green dashed-dotted line differs from the first one by taking into account the nontrivial QED contribution in zeroth order in $1/Z$. The violet dotted line is obtained by adding the higher-order (in $\alpha Z$) contribution in first order in $1/Z$. Finally, the red solid line includes also the finite nuclear size correction $\delta A_{\rm Breit}^{\rm fns,1el}(\alpha Z)$ and corresponds to the total data presented in Table~\ref{tab:total}. We note that the last two corrections have a different sign and partly cancel each other in the sum. These corrections have not been taken into account in Ref.~\cite{Zubova:2014:062512}. The Breit-approximation values and the results with the QED contribution evaluated within the independent-electron approximation from Ref.~\cite{Zubova:2014:062512} are shown in Fig.~\ref{fig:step_by_step:tr_2p1_2s} with the magenta circles and black diamonds, respectively. In order not to overload the plot, we omit the error bars in Fig.~\ref{fig:step_by_step:tr_2p1_2s}(a). The uncertainties are indicated only in Fig.~\ref{fig:step_by_step:tr_2p1_2s}(b). One can see that there is a reasonable agreement between the data from Ref.~\cite{Zubova:2014:062512} and the results of the present study. The difference between the final theoretical predictions is explained by the fact that the more subtle effects are taken into account now. As a result, the uncertainty of the nuclear recoil effect is reduced, especially for middle-$Z$ ions, where the contribution of the mass shift to the isotope shifts is more significant. The results obtained are in demand in view of the existing and forthcoming experimental investigations of the relativistic and QED nuclear recoil effect~\cite{Brandau:2003:073202, SoriaOrts:2006:103002, Brandau:2008:073201, Brandau:2013:014050}.


\section{Summary \label{sec:4}}

To summarize, we have developed the rigorous QED formalism which allows us to calculate the electron-electron interaction correction to the one-electron part of the nuclear recoil effect on binding energies in atoms and ions nonperturbatively in the parameter $\alpha Z$. The method derived was employed for the \textit{ab initio} calculations of the one-electron nuclear recoil contribution to the binding energies of the $1s^2$ state in He-like ions and $1s^2 2s$ and $1s^2 2p_{1/2}$ states in Li-like ions in the wide range $Z=5$--$100$. The corresponding contribution to the $2p_{1/2}$--$2s$ transition energy in Li-like ions was studied as well. The one-electron part of the nuclear recoil effect was combined with the two-electron part considered recently in Ref.~\cite{Malyshev:2019:012510}. The all-order (in $\alpha Z$) results to zeroth and first orders in $1/Z$ were compared with the values obtained by applying the mass shift operator $H_{M}$. The nontrivial QED contribution was extracted, and its behavior with the growth of $Z$ was investigated. This provides an estimation of the accuracy of the calculations based on the mass shift operator which is valid within the $(m/M)(\alpha Z)^4 mc^2$ approximation only. Finally, the QED calculations to first order in $1/Z$ were supplemented with the higher-order correlation corrections evaluated within the Breit approximation. As a result, the most accurate theoretical predictions for the mass shifts of the binding and transition energies in He- and Li-like ions have been obtained.


\section*{Acknowledgments}

We thank Ilya Tupitsyn for valuable discussions. The work was supported by RFBR (Grant No.~18-32-00294). A.V.M., M.Y.K., and V.M.S. acknowledge the support from the Foundation for the advancement of theoretical physics and mathematics ``BASIS''. 




\end{document}